\documentstyle[11pt]{article}
\normalbaselines

\newtheorem{thm}{Theorem}[section]
\newtheorem{lem}{Lemma}[section]

\begin{document}

\def\bea*{\begin{eqnarray*}}
\def\eea*{\end{eqnarray*}}
\def\ba{\begin{array}}
\def\ea{\end{array}}
% --------------------------------------------------------------
% If you want to see the names of the equations and references
% set below \count1=0 otherwise \count1=1
% --------------------------------------------------------------
\count1=1
% --------------------------------------------------------------
\def\be{\ifnum \count1=0 $$ \else \begin{equation}\fi}
\def\ee{\ifnum\count1=0 $$ \else \end{equation}\fi}
\def\ele(#1){\ifnum\count1=0 \eqno({\bf #1}) $$ \else \label{#1}\end{equation}\fi}
\def\req(#1){\ifnum\count1=0 {\bf #1}\else \ref{#1}\fi}
\def\bea(#1){\ifnum \count1=0   $$ \begin{array}{#1}
\else \begin{equation} \begin{array}{#1} \fi}
\def\eea{\ifnum \count1=0 \end{array} $$
\else  \end{array}\end{equation}\fi}
\def\elea(#1){\ifnum \count1=0 \end{array}\label{#1}\eqno({\bf #1}) $$
\else\end{array}\label{#1}\end{equation}\fi}
\def\cit(#1){
\ifnum\count1=0 {\bf #1} \cite{#1} \else 
\cite{#1}\fi}
\def\bibit(#1){\ifnum\count1=0 \bibitem{#1} [#1    ] \else \bibitem{#1}\fi}
\def\ds{\displaystyle}
\def\hb{\hfill\break}
\def\comment#1{\hb {***** {\em #1} *****}\hb }

\newcommand{\TZ}{\hbox{\bf T}}
\newcommand{\MZ}{\hbox{\bf M}}
\newcommand{\ZZ}{\hbox{\bf Z}}
\newcommand{\NZ}{\hbox{\bf N}}
\newcommand{\RZ}{\hbox{\bf R}}
\newcommand{\CZ}{\,\hbox{\bf C}}
\newcommand{\PZ}{\hbox{\bf P}}
\newcommand{\QZ}{\hbox{\rm eight}}
\newcommand{\HZ}{\hbox{\bf H}}
\newcommand{\EZ}{\hbox{\bf E}}
\newcommand{\GZ}{\,\hbox{\bf G}}

\font\germ=eufm10
\def\goth#1{\hbox{\germ #1}}
\vbox{\vspace{38mm}}

\begin{center}
{\LARGE \bf The Transfer Matrix of Superintegrable Chiral Potts Model as the Q-operator of  Root-of-unity XXZ Chain with Cyclic Representation of ${\bf U_{\sf q}(sl_2)}$ } \\[10 mm] 
Shi-shyr Roan \\
{\it Institute of Mathematics \\
Academia Sinica \\  Taipei , Taiwan \\
(email: maroan@gate.sinica.edu.tw ) } \\[25mm]
\end{center}

\begin{abstract}
We demonstrate that the transfer matrix of the inhomogeneous $N$-state chiral Potts model with two vertical superintegrable rapidities serves as the $Q$-operator of XXZ chain model for a cyclic representation of $U_{\sf q}(sl_2)$ with $N$th root-of-unity ${\sf q}$ and representation-parameter for odd $N$. The symmetry problem of XXZ chain with a general cyclic $U_{\sf q}(sl_2)$-representation is mapped onto the problem of studying $Q$-operator of some special one-parameter family of generalized $\tau^{(2)}$-models. In particular, the spin-$\frac{N-1}{2}$ XXZ  chain model with ${\sf q}^N=1$ and the homogeneous $N$-state chiral Potts model at a specific superintegrable point are unified as one physical theory. By Baxter's method developed for producing $Q_{72}$-operator of the root-of-unity eight-vertex model, we construct the $Q_R, Q_L$- and $Q$-operators of a superintegrable $\tau^{(2)}$-model, then identify them with transfer matrices of the $N$-state chiral Potts model for a positive integer $N$. We thus obtain a new method of producing the superintegrable $N$-state chiral Potts transfer matrix from the $\tau^{(2)}$-model by constructing its $Q$-operator.

\end{abstract}
\par \vspace{5mm} \noindent
{\rm 2006 PACS}:  05.50.+q, 02.20.Uw, 75.10Jm \par \noindent
{\rm 2000 MSC}: 14H70, 39B72, 82B23  \par \noindent
{\it Key words}: $\tau^{(2)}$-model, Chiral Potts model, Q-operator, Quantum algebra $U_{\sf q}(sl_2)$  \\[10 mm]

\setcounter{section}{0}
\section{Introduction}
\setcounter{equation}{0}
The aim of this paper is to show that the transfer matrix of some inhomogeneous $N$-state chiral Potts model\footnote{All the models discussed in this paper always assume with the periodic condition.} (CPM) with two vertical superintegrable rapidities serves as the $Q$-operator of the 
XXZ chain model arisen from cyclic representations of $U_{\sf q}(sl_2)$ with ${\sf q}^N= \varsigma^N =1$ for odd $N$, where $\varsigma= {\sf q}^{2\varepsilon+2}$ with $\varepsilon$   as the parameter of $N$-dimensional $U_{\sf q}(sl_2)$-cyclic representations (see (\req(crep)) in the paper). In particular, the cyclic representation for $\varsigma = {\sf q}$ is equivalent to the spin-$\frac{N-1}{2}$ representation of $U_{\sf q}(sl_2)$.  
As a consequence, the result yields the identical physical theory about the spin-$\frac{N-1}{2}$ XXZ  chain with ${\sf q}^N=1$, and the homogeneous $N$-state CPM at some specific superintegrable point. In recent years, the degeneracy of the spin-$\frac{1}{2}$ XXZ Hamiltonian with the extra $sl_2$-loop-algebra symmetry was discovered and analysed in \cite{DFM, De05, FM00, FM001, FM01} when the (anisotropic) parameter ${\sf q}$ is a root of unity. The $sl_2$-loop-algebra symmetry was further verified in the spin-$\frac{d-1}{2}$ XXZ chain with ${\sf q}^N=1$ for $2 \leq d \leq N$ when $N$ is odd \cite{NiD, R06F}. In the present work, we study the spin-$\frac{N-1}{2}$ XXZ chain with ${\sf q}^N=1$ restricted only in the odd $N$ case, which for convenience, will be loosely called the root-of-unity XXZ chain with ${\sf q}^N=1$ throughtout the paper. The scheme of our investigaton is to build an explicit connection between the theories of root-of-unity XXZ chains with ${\sf q}^N=1$ and superintegrable $N$-state CPM through the $Q$-operator approach, while discussions in CPM and its related $\tau^{(2)}$-model will be understandingly valid for all integers $N \geq 2$ with no oddness restriction. Even though the spin-$\frac{N-1}{2}$ XXZ  chain with ${\sf q}^N=1$ and
the homogeneous $N$-state CPM share the same Bethe equation (up to phrase factors) \cite{AMP, B93, DKM}, the symmetry structures of these two models are different (though closely related):
the  superintegrable CPM with Onsager-algebra symmetry \cite{R05o}, while the spin-$\frac{N-1}{2}$ XXZ chain with ${\sf q}^N=1$ carries the $sl_2$-loop-algebra symmetry. However the subtle correspondence made in \cite{R06F} about the symmetry comparison by the functional-relation method has strongly suggested that one model could possess some additional structure which already appeared in the other one. But the conclusion was not available within one theory alone. 
In the present paper we show that the connection indeed is established for the conjectural relationship about the identical theory of superintegrable CPM and the spin-$\frac{N-1}{2}$ XXZ chain with ${\sf q}^N=1$.  
The identification of these two models, not merely on the existence of quantitative analogy between them, can provide deeper insights about natures of these theories, e.g., the extension of the Onsager-algebra symmetry of superintegrable CPM to the larger $sl_2$-loop-algebra symmetry; a similar symmetry description also true for the spin-$\frac{N-1}{2}$ XXZ chain.
The finding has now further demonstrated the universal role of CPM  about the symmetry structure of various lattice models, such as a not yet complete task of the root-of-unity XXZ spin chains and eight-vertex model in \cite{DFM, De01a, De01b, FM01, FM02, FM04,FM41, F06, NiD, R05b, R06Q, R06F, R06Q8, R07}. Furthermore, the oddness restriction of $N$ for the root-of-unity XXZ chains in this paper is purely imposed by the technique requirement in the mathematical discussion, but not a conceptual one. Hence the relation between $sl_2$-loop-algebra symmetry of spin-$\frac{N-1}{2}$ XXZ chain and Onsager-algebra symmetry of $N$-state CPM could still be valid for even $N$ if the computational techniques can be extended (though not immediately apparent at present) to the general case.

The chiral Potts model was originally presented as an $N$-state one-dimensional quantum Hamiltonian \cite{HKN, GR}, a formulation implanting the character of Onsager-algebra symmetry in the theory. Then it was formulated as a two-dimensional solvable lattice model in statistical mechanics which satisfies the star-triangle relations \cite{AMPTY, MPTS, BPA}. For $N=2$, it  reduces the Ising model, of which the free energy was solved by Onsager in 1944 \cite{On}. When $N \geq 3$, due to the lack of difference property of rapidities, a characteristic feature in the study of CPM, e.g. the eigenvalues \cite{B90, B93, MR} and the order parameter \cite{B05o, B05}, relies on functional relations among the CPM and various related $\tau^{(j)}$-models \cite{BBP}. This method stemmed from the descendent relation of CPM with the six-vertex model in \cite{BazS}, where there exhibited a five-parameter Yang-Baxter solutions for the asymmetric six-vertex $R$-matrix (see (\req(ttL)) in this paper), called the generalized $\tau^{(2)}$-model, among which are the $\tau^{(2)}$-matrices parametrized by rapidities of CPM. In this work, we study the XXZ chains associated to cyclic representations of $U_{\sf q}(sl_2)$, and find that they are equivalent to some special one-parameter family of generalized $\tau^{(2)}$-model, but not in common with $\tau^{(2)}$-matrices for the homogenous CPM except one superintegrable point. Accordingly, the symmetry study of those  
XXZ chains is thus mapped onto the functional-relation study of the corresponding $\tau^{(2)}$-models, hence an appropriate $Q$-operator is required for this purpose. 
In this paper, we employ the Baxter's techniques developed for producing $Q_{72}$-operator of the root-of-unity eight-vertex model \cite{B72} to construct the $Q_R, Q_L$-, then $Q$-operators for a $\tau^{(2)}$-model corresponding to the XXZ chain for a root-of-unity $U_{\sf q}(sl_2)$-cyclic representation, much in the same way as the $Q$-operator of the root-of-unity six-vertex model in \cite{R06Q}. The method is successfully applied  to the superintegrable $\tau^{(2)}$-matrix so that the homogeneous CPM transfer matrices $T_p, \widehat{T}_p$ \cite{BBP} at an arbitrary superintegrable element $p$ are recognized as the $Q_R, Q_L$-operators (up to certain normalized factors) by a correct identification of various parameters appeared in the construction. Also, the procedure provides the reasoning for the high-genus-curve constraint of the rapidity. We thus obtain another $Q$-operator method, (implicitly related to arguments in \cite{BBP, BazS}), of creating the superintegrable CPM transfer matrix from the $\tau^{(2)}$-matrix. The same procedure enables us to construct the $Q_R, Q_L$- and $Q$-operators of the $\tau^{(2)}$-matrix corresponding to a cyclic representation of $U_{\sf q}(sl_2)$ with a $N$th root-of-unity representation-parameter, then identify them with the inhomogeneous CPM transfer matrices with two superintegrable vertical rapidities. 
The inhomogeneity of CPM may suggest the possible significant role of $Q$-operator in the study of generalized $\tau^{(2)}$-models.

This paper is organized as follows. In section \ref{sec:CPtau}, we briefly review the rapidity  and the transfer matrix of the $N$-state CPM in \cite{BBP},  and main features of the generalized $\tau^{(2)}$-model related to the superintegrable CPM in \cite{BazS} (or   \cite{R05o} and references therein). In section \ref{sec:Qtau}, we reproduce the homogenous CPM transfer matrix as the $Q$-operator of the superintegrable $\tau^{(2)}$-matrix along the line of Baxter's $Q_{72}$-method in \cite{B72}. We construct  the $Q_R, Q_L$-, and  $Q$-operators of a superintegrable $\tau^{(2)}$-matrix, first in some detail for the special superintegrable element (\req(sup)) in section \ref{ssec.Qsup1}, then at a general superintegrable element (\req(supp)) in section \ref{ssec.QSUP}. 
The rapidity constraint of the high-genus curve in CPM is revealed as a requirement for the commutative property of  $Q$-matrices. In section \ref{sec:QXXZ}, we study the $Q$-operator of the XXZ chain model associated to cyclic representations of $U_{\sf q}(sl_2)$ with ${\sf q}$ a $N$th root of unity, among which the spin-$\frac{N-1}{2}$ highest-weight representation appears as a special case for odd $N$. First in section \ref{ssec.cyXXZ}, we illustrate that the XXZ chain model for cyclic representations of $U_{\sf q}(sl_2)$ are equivalent to a special one-parameter family of generalized $\tau^{(2)}$-model. In particular, the result yields the identical theory of  the spin-$\frac{N-1}{2}$ XXZ  chain and  the homogeneous $N$-state CPM at one specific superintegrable point. Then in section \ref{ssec.IcpQ}, we construct the $Q$-operator of the  special  generalized $\tau^{(2)}$-model with $N$th root-of-unity parameters, equivalently the XXZ chain for those cyclic $U_{\sf q}(sl_2)$-representations. We then identify these $Q$-operators with the transfer matrix of the inhomogeneous $N$-state CPM with certain two vertical superintegrable rapidities. 
We close in section \ref{sec. F} with some concluding remarks.

\section{The $N$-state Chiral Potts Model and the Generalized $\tau^{(2)}$-model \label{sec:CPtau}}
\setcounter{equation}{0}
This section serves as a brief introduction to the chiral Potts model and the generalized $\tau^{(2)}$-model. 
The summary will be sketchy, but also serve to establish the notation (for more details, see \cite{AMP, BBP, BazS, R05o} and references therein).

In this paper, $\CZ^N$ denotes the vector space consisting of $N$-cyclic vectors $v = \sum_{n \in \ZZ_N} v_n | n \rangle$ with the basis indexed by  $n \in \ZZ_N (:= \ZZ/N\ZZ)$. We fix the $N$th root of unity $\omega = e^{\frac{2 \pi \sqrt{-1}}{N}}$, and a pair of Weyl $\CZ^N$-operators, $X$ and  $Z$, with the relations $XZ= \omega^{-1}ZX$ and  $X^N=Z^N=1$:
$$
 X |n \rangle = | n +1 \rangle , ~ \ ~ Z |n \rangle = \omega^n |n \rangle ~ ~ \ ~ ~ (n \in \ZZ_N) .
$$

The rapidities  of the $N$-state CMP are described by coordinates $(x , y , \mu ) \in \CZ^3$  satisfying the following equations of a genus-$(N^3-2N^2+1)$ curve  
\be
{\goth W}_{k'}: ~ ~ k x^N  = 1 -  k'\mu^{-N}, \ ~ \  k  y^N  = 1 -  k'\mu^N  \ ,
\ele(xymu)
where $k' (\neq \pm 1, 0)$ is the temperature-like parameter with $k = \sqrt{1- k'^2}$.  The elements in ${\goth W}_{k'}$ will be denoted by $p, q, \cdots$, and its coordinates will be written by $x_p, y_p, \mu_p$ whenever if it will be necessary to specify the element $p$. Denote $t_p= x_p y_p$. 
The Boltzmann weights of the $N$-state CPM are defined by coordinates of $p, q \in {\goth W}_{k'}$ with the expressions: 
\be
\frac{W_{p,q}(n)}{W_{p,q}(0)}  = (\frac{\mu_p}{\mu_q})^n \prod_{j=1}^n
\frac{y_q-\omega^j x_p}{y_p- \omega^j x_q }, ~ \
\frac{\overline{W}_{p,q}(n)}{\overline{W}_{p,q}(0)}  = ( \mu_p\mu_q)^n \prod_{j=1}^n \frac{\omega x_p - \omega^j x_q }{ y_q- \omega^j y_p }.
\ele(CPW)
The rapidity constraint (\req(xymu)) ensures the above Boltzmann weights with the $N$-periodicity property for $n$. Then the star-triangle relation holds:
\be
\sum_{n=0}^{N-1} \overline{W}_{qr}(j' - n) W_{pr}(j - n) \overline{W}_{pq}(n - j'')= R_{pqr} W_{pq}(j - j')\overline{W}_{pr}(j' - j'') W_{qr}(j -j'')    
\ele(TArel)
where the factor $R_{pqr}$ is defined by
\be
R_{pqr}= \frac{f_{pq}f_{qr}}{f_{pr}}, \ \ f_{pq} : = \bigg( \frac{{\rm det}_N( \overline{W}_{pq}(i-j))}{\prod_{n=0}^{N-1} W_{pq}(n)}\bigg)^{\frac{1}{N}}.
\ele(Rf) 
Without loss of generality, we may assume $W_{p, q}(0) = \overline{W}_{p,q}(0) = 1$.
On a lattice of the horizontal size $L$, the combined weights of
intersection between two consecutive rows give rise
to an operator of $\stackrel{L}{\otimes} \CZ^N$, which defines the transfer matrix of the $N$-state CPM\footnote{We use the convention of transfer matrices in \cite{BBP} where formulas (2.15a) (2.15b) are the (\ref{Tpq}) (\ref{hTpq}) here. The transfer matrix  \cite{AMP}(1.6)  or  \cite{R05o} (7)   is equal to (\ref{hTpq}) in this paper.} :
\begin{eqnarray}
T_p (q)_{\{j \}, \{j'\}} = \prod_{\ell =1}^L W_{p,q}(j_\ell - j'_\ell )
\overline{W}_{p,q}(j_{\ell+1} - j'_\ell),
\label{Tpq}
\end{eqnarray}
for $p, q \in {\goth W}_{k'}$ and  $j_\ell, j'_\ell \in \ZZ_N$. Here the periodic condition is imposed by defining $L+1=1$, hence $T_p( q)$ commutes with the spatial translation
\be
S_R : | j_1, \ldots, j_L \rangle \mapsto | j_2,  \ldots, j_{L+1} \rangle ~ ~ ~ j_\ell \in \ZZ_N . 
\ele(SR)
The operator $\widehat{T}_p = T_p S_R$ is expressed by 
\begin{eqnarray}
\widehat{T}_p (q)_{\{j \}, \{j'\}} = \prod_{\ell =1}^L \overline{W}_{p,q}(j_\ell - j'_\ell) W_{p,q}(j_\ell - j'_{\ell+1}).
\label{hTpq}
\end{eqnarray}
The star-triangle relation (\req(TArel)) in turn yields the commuting 
transfer matrices for a fixed $p \in {\goth W}_{k'} $:
$$
[ T_p (q) , \ T_p (q') ]= [ \widehat{T}_p (q) , \ \widehat{T}_p (q') ] = 0 \  \ , \ \ \ q , q' \in {\goth W}_{k'} \ . 
$$

In the discussion of  CPM as a descendent of the six-vertex model in \cite{BazS}, 
a five-parameter family of generalized $\tau^{(2)}$-models was discovered as the
Yang-Baxter (YB) solution for the {\it asymmetric} six-vertex $R$-matrix,
$$
R(t) = \left( \begin{array}{cccc}
        t \omega - 1  & 0 & 0 & 0 \\
        0 &t-1 & \omega  - 1 &  0 \\ 
        0 & t(\omega  - 1) &( t-1)\omega & 0 \\
     0 & 0 &0 & t \omega - 1    
\end{array} \right).
$$
The $L$-operator of those $\tau^{(2)}$-models  
is a matrix of $\CZ^2$-auxiliary and  $\CZ^N$-quantum space built upon the Weyl operators $X, Z$ with parameters $\alpha, \beta, \gamma, \varrho, \kappa \in \CZ$:
\be
{\tt L} ( t ) =  \left( \begin{array}{cc}
        1 + t \kappa X  & ( \gamma - \varrho X)Z \\
        t ( \alpha - \beta X)Z^{-1} & t \alpha \gamma + \frac{\beta \varrho}{\kappa} X 
\end{array} \right)=: \left( \begin{array}{cc}
        {\sf A}  & {\sf B}   \\
        {\sf C} & {\sf D}  
\end{array} \right) (t), \ \ t \in \CZ ,
\ele(ttL) 
which satisfy the YB equation
\be
R(t/t') ({\tt L} (t) \bigotimes_{aux}1) ( 1
\bigotimes_{aux} {\tt L} (t')) = (1
\bigotimes_{aux} {\tt L}(t'))( {\tt L}(t)
\bigotimes_{aux} 1) R(t/t').
\ele(YB)
Then the monodromy matrix of size $L$, 
$$
\bigotimes_{\ell=1}^L  {\tt L}_\ell (t) = \left( \begin{array}{cc}
        A(t)  & B(t) \\
        C(t) & D(t)
\end{array} \right), ~ ~ ~ {\tt L}_\ell = {\tt L} ~ ~ {\rm at ~ site} ~ \ell ,
$$
again satisfies the YB equation, and the $\omega$-twisted trace
\be
\tau^{(2)}(t) = A(\omega t) + D(\omega t) ,
\ele(tau2)
form a family of commuting operators of the $L$-tensor space $\stackrel{L}{\otimes} \CZ^N$ of $\CZ^N$. By 
\be
{[X , A ]} =  [X , D ] = 0 , \ \ X B = \omega^{-1} B X , \ \ X C = \omega C X ,
\ele(XAB)
$X$ commutes with the $\tau^{(2)}$-matrix.
The quantum determinant of the monodromy matrix is characterized by rank-one property of 
$R( \omega^{-1})$ in the YB relation (\req(YB)):
$$
\begin{array}{l}
R(\omega^{-1}) (\otimes {\tt L}_\ell (t) \bigotimes_{aux}1) ( 1
\bigotimes_{aux} \otimes {\tt L}_\ell (\omega t)) = \\ (1 \bigotimes_{aux} \otimes {\tt L}_\ell (\omega t))( \otimes {\tt L}_\ell (t)
\bigotimes_{aux} 1) R(\omega^{-1}) 
= {\rm det}_q \cdot  R(\omega^{-1}),
\end{array}
$$
with ${\rm det}_q (={\rm det}_q (\otimes {\tt L}_\ell) (t)) = (\frac{\beta \varrho}{\kappa} + (\alpha \varrho + \omega \beta \gamma ) t + \omega \alpha \gamma \kappa t^2)^L X^L $; equivalently, the following relations hold:
\bea(rll)
& B(\omega t) A (t) = A(\omega t) B (t), & D(\omega t) C (t) = C (\omega t) D (t); \\ 
&A( t) C (\omega t) = \omega C(t) A (\omega t),& B(t) D(\omega t) = \omega D(t) B(\omega t), \\
{\rm det}_q  & = D(\omega t) A(t) - C(\omega t) B(t) &= A(\omega t) D(t) - B(\omega t) C(t)  \\ 
& = A(t)D(\omega t) - \omega C(t) B(\omega t) &= D(t)A(\omega t) - \omega^{-1}  B(t) C(\omega t) . 
\elea(YBc)
For the $\tau^{(2)}$-matrix in CPM with the rapidity $p=(x, y, \mu)$ in (\req(xymu)), the parameters in (\req(ttL)) are set by  
\be
\alpha = -\gamma = - y^{-1} , ~ \beta = - \omega^{-1} \varrho = \frac{-x\mu^2 }{ y^2}, ~ \kappa = \frac{-\mu^2}{y^2} .  
\ele(cpmL)
Hence the parameters (\req(ttL)) for the superintegrable point in ${\goth W}_{k'}$: 
\be
x_p = y_p = \eta^{\frac{1}{2}} , ~ ~ \mu_p = 1 , ~ ~  {\rm where} ~ \eta:= (\frac{1-k'}{1+k'})^{\frac{1}{N}}, 
\ele(sup)
are given by $
-\alpha = -\beta = \gamma  =  \omega^{-1} \varrho =  - \kappa  =  \eta^{\frac{-1}{2}}$.
It is known that the degeneracy of $\tau^{(2)}$-eigenvalues with Onsager-algebra symmetry occurs at the superintegrable point (\req(sup)) \cite{R04, R05o}.  
With the variable ${\tt t} = \eta^{-1} t$, the superintegrable $L$-operators for an arbitrary $k'$ is gauge equivalent to the one with $\eta=1$ (\cite{R05o} section 4):
\be
{\tt L} ( {\tt t} ) =  \left( \begin{array}{cc}
        1 - {\tt t} X  & ( 1 - \omega X)Z \\
        - {\tt t} ( 1 - X)Z^{-1} & - {\tt t}  +  \omega X 
\end{array} \right) =: \left( \begin{array}{cc}
        {\sf A}  & {\sf B}   \\
        {\sf C} & {\sf D}  
\end{array} \right) ({\tt t}) .
\ele(SupL) 
In the next section, we produce a $Q$-operator of the above $L$-operator following Baxter's method of constructing $Q_{72}$-operator for the eight-vertex model in \cite{B72}. The constructed $Q$-operator will be identified with the CPM transfer matrix (\ref{Tpq}) at the superintegrable point (\req(sup)) in a transparent manner.

\section{The $Q$-operator of Superintegrable $\tau^{(2)}$-model and CPM Transfer Matrix  \label{sec:Qtau}}
\setcounter{equation}{0}
In this section by Baxter's method of producing $Q_{72}$-operator in \cite{B72}, we construct the $Q_R, Q_L$-, then $Q$-operators of a homogeneous superintegrable $\tau^{(2)}$-model, first for the $L$-operator (\req(SupL)) in section \ref{ssec.Qsup1}, then for a general superintegrable $L$-operator (\req(SUL)) in section \ref{ssec.QSUP}. 
In doing so, we reproduce the CPM transfer matrix \cite{BBP} at an arbitrary superintegrable point. 

\subsection{Derivation of the CPM transfer matrix as the $Q$-operator of the superintegrable $\tau^{(2)}$-model  \label{ssec.Qsup1}}
The $Q_R$-matrix associated to the $L$-operator  (\req(SupL)) is constructed from an ${\sf S }$-operator, which is a matrix of the $\CZ^N$-auxiliary and $\CZ^N$-quantum space 
\be
{\sf S }  = ({\sf S }_{i,j})_{i, j \in \ZZ_N}  
\ele(Smat)
with $\CZ^N$-operator entries ${\sf S }_{i,j}$. The $Q_R$-operator is defined by 
\be
Q_R= {\rm tr}_{\CZ^N} ( \bigotimes_{\ell =1}^L {\sf S}_{\ell}), \ \ \ {\sf S}_{\ell}= {\sf S} \ {\rm at \ site} \ \ell,
\ele(QR)
by which, $\tau^{(2)} Q_R = {\rm tr}_{\CZ^2 \otimes \CZ^N} ( \bigotimes_{\ell =1}^L {\sf U}_{\ell})$, where ${\sf U}_{\ell}= {\sf U}$ at the site $\ell$ with the local-operator ${\sf U}$ being a matrix of $\CZ^2 \otimes \CZ^N $-auxiliary and $\CZ^N$-quantum space:
$$
{\sf U} = \left( \begin{array}{cc}
        {\sf A} {\sf S } & {\sf B} {\sf S} \\
        {\sf C} {\sf S} & {\sf D} {\sf S }
\end{array} \right) .
$$
Hereafter we write the operators ${\sf A} (t), {\sf B} (t), {\sf C}(t), {\sf D}(t) $ simply by ${\sf A}, {\sf B}, {\sf C}, {\sf D}$ if no confusion could arise; while the matrix ${\sf S}$ will depend on some variable $\sigma$ algebraically related to the variable $t$: ${\sf S} = {\sf S}(\sigma)$. 
The operator $\tau^{(2)}Q_R$ will be decomposed into the sum of two matrices if we can find a $2N$ by $2N$ scalar matrix (independent of $s$) 
\be
{\sf M} = \left( \begin{array}{cc}
        I_N  & 0 \\
       -\delta & I_N 
\end{array} \right) , \ \ \delta = {\rm dia} [\delta_0, \cdots, \delta_{N-1}] ,
\ele(Md)
so that $
{\sf M}^{-1} {\sf U} {\sf M} = \left( \begin{array}{cc}
        * & * \\ 
         0 & *
\end{array} \right)$. One can express ${\tt M}^{-1} {\tt U} {\tt M}$ by\footnote{The ${\tt A}_{\delta_{j}}, {\tt C}_{\delta_i, \delta_{j}}, {\tt D}_{\delta_i}$ here are  ${\tt A}(-\delta_{j}), {\tt C}(-\delta_i, -\delta_{j}), {\tt D}(-\delta_i)$ in \cite{R07} (3.2).}
$$
{\tt M}^{-1} {\tt U} {\tt M} = \left( \begin{array}{cc}
        {\tt A}_{\delta_j}{\tt S }_{i,j}  ,  & {\tt B} {\tt S }_{i,j}   \\ 
       {\tt C}_{\delta_i, \delta_{j}}{\tt S }_{i,j}  , & {\tt D}_{\delta_i} {\tt S }_{i,j} 
\end{array} \right)_{i, j \in \ZZ_N} .
$$
Here the $\CZ^N$-operators ${\tt A}_{\eta}, {\tt C}_{\xi, \eta}, {\tt D}_{\xi}$ for $\xi, \eta \in \CZ$ associated to a $L$-operator in (\req(ttL)) are defined by
\bea(l)
{\tt A}_{\eta}(t):= {\tt A}(t)  -  \eta {\tt B}(t),  ~ ~ ~  {\tt D}_{\xi}(t):= \xi {\tt B}(t)   + {\tt D}(t)  ,   \\
{\tt C}_{\xi, \eta} (t) :=  \xi {\tt A}(t)  + {\tt C}(t) - \xi \eta {\tt B}(t)  -{\tt D}(t)\eta .
\elea(gauL)
\begin{lem}\label{lem:gADC} 
The operators ${\tt A}_{\eta}, {\tt C}_{\xi, \eta}, {\tt D}_{\xi}$~$(\xi, \eta \in \CZ)$ in $(\req(gauL))$ associated to a $L$-operator in $(\req(ttL))$ satisfy the following commutative relations:
\bea(l)
{\tt C}_{\xi, \eta} (t) X^{-1} {\tt A}_{\eta}(\omega t) =  {\tt A}_{\eta} (t)X^{-1}{\tt C}_{\xi, \eta}(\omega t), \\
{\tt C}_{\xi, \eta} (\omega t) {\tt D}_{\xi}(t) =  {\tt D}_{\xi} (\omega t) {\tt C}_{\xi, \eta}(t) .
\elea(ADCg)
\end{lem}
{\it Proof.} With the expression, $
{\tt C}_{\xi, \eta} =  \xi {\tt A}_{\eta} + {\tt C} - {\tt D} \eta =  \xi {\tt A}  + {\tt C}  -{\tt D}_{\xi} \eta$, and relations, (\req(XAB)) and 
(\req(YBc)), one finds 
$$
\begin{array}{ll}
({\tt C} - {\tt D} \eta)(t)X^{-1} {\tt A}_{\eta}(\omega t) &= {\tt A}_{\eta}(t)X^{-1} ({\tt C}- {\tt D} \eta)(\omega t), \\
( \xi {\tt A}  + {\tt C})(\omega t) {\tt D}_{\xi}(t) &= {\tt D}_{\xi}(\omega t) (\xi {\tt A}   + {\tt C})(t).
\end{array}
$$
Then follows (\req(ADCg)).
 \par \vspace{.1in}

Now we determine the condition of $\xi, \eta$ with the singular matrix ${\tt C}_{\xi, \eta}$ for the $L$-operator in (\req(SupL)). 
\begin{lem}\label{lem:Ckck} 
The criterion for $\xi, \eta$ with a singular matrix ${\tt C}_{\xi, \eta}$ associated to the $L$-operator $(\req(SupL))$ is $\xi^N = \eta^N$. When $\eta = \omega^{-k} \xi $,  the kernel space of ${\tt C}_{\xi, \eta}$ is one-dimensional with the cyclic-vector basis $v = \sum_{n \in \ZZ_N} v_n |n \rangle \in \CZ^N$ expressed by
\be
\frac{v_n}{v_{n-1}} = \frac{( \omega -\omega^n \xi  ) ({\tt t} -  \omega^{n-k} \xi)}{(1 -  \omega^{n-k}\xi ) ( {\tt t}- \omega^n \xi )} \ ~ \ ( n \in \ZZ_N) 
\ele(v)
satisfying the relations
\bea(l)
{\tt A}_{\eta}({\tt t})v({\tt t}) = (1 -  \omega^{-1} {\tt t} )   \frac{   ({\tt t}- \xi \omega^{-k+1}    )  v({\tt t})_0}{(  \omega^{-1} {\tt t} -  \xi   )  v (\omega^{-1} {\tt t})_0}  X v (\omega^{-1} {\tt t}) , \\ {\tt D}_{\xi}({\tt t})v({\tt t}) = \omega (1-{\tt t} )  \frac{ ( {\tt t} -\xi ) v({\tt t})_0}{ (\omega {\tt t}- \xi \omega^{1-k}   )v (\omega {\tt t})_0} v (\omega {\tt t}) .
\elea(ADv)
Similarly when $\eta' = \omega^{-k} \xi' $, the cokernel of ${\tt C}_{\xi', \eta'}$ is the one-dimensional space with the basis element $v^* = \sum_{n \in \ZZ_N} v^{* n} \langle n | \in \CZ^{N *}$ expressed by
\be
\frac{v^{* n}}{v^{* n-1}} = \frac{ (1 -  \omega^{n-k-1} \xi')({\tt t} - \omega^{n-1} \xi'  )}{ ( 1- \omega^{n-1} \xi'  ) ({\tt t} -  \omega^{n-k} \xi')}  \ ~ \ ( n \in \ZZ_N) 
\ele(v*)
satisfying the relations
\bea(l)
v^*({\tt t}) {\tt A}_{\eta'} ({\tt t})  = (1-{\tt t}) \frac{ (\omega^{-k} \xi' - {\tt t})  v^*({\tt t})^0 }{ (\omega ^{-1} \xi' - {\tt t} ) v^*(\omega {\tt t})^0 } v^*(\omega {\tt t}) X , \\
v^*({\tt t}) {\tt D}_{\xi'} ({\tt t}) = (\omega -{\tt t}) \frac{ (\xi' - {\tt t}) v^*({\tt t})^0 }{  (\omega^{-k+1} \xi' - {\tt t}) v^*(\omega^{-1} {\tt t})^0 } v^*(\omega^{-1} {\tt t}) .
\elea(v*AD)
\end{lem}
{\it Proof.} By (\req(SupL)), the entries of ${\tt C}_{\xi, \eta}$ are zeros except
$$
\langle  n | {\tt C}_{\xi, \eta}| n \rangle = (\xi - \omega^{-n} {\tt t} ) (1 - \eta \omega^n) , ~ \ ~  \langle  n | {\tt C}_{\xi, \eta}| n-1 \rangle = -( \xi - \omega^{-n+1} ) ({\tt t} - \eta \omega^n) 
$$
for $n=0, \ldots , N-1 \in \ZZ_N$. The kernel vector $v=\sum_{n \in \ZZ_N} v_n |n \rangle$ and covector $v^* =\sum_{n \in \ZZ_N} v^{* n} \langle n |$ of ${\tt C}_{\xi, \eta}$ are determined by the relations
$$
\begin{array}{ll}
( \xi - \omega^{-n+1} ) ({\tt t} - \eta \omega^n) v_{n-1} &= ( \xi - \omega^{-n} {\tt t} ) (1 - \eta \omega^n) v_n , \\
(\xi - \omega^{-n+1} {\tt t} ) (1 - \eta \omega^{n-1}) v^{* n-1} &= ( \xi - \omega^{-n+1} ) ({\tt t} - \eta \omega^n)  v^{* n}  ,
\end{array}
$$
for $n=0, \ldots , N-1$. The non-zero vector condition for $v$ is given by $v_0 = v_N \neq 0$, equivalently, $
\prod_{n=0}^{N-1} \frac{( \xi - \omega^{-n+1} ) ({\tt t} - \eta \omega^n)}{( \xi - \omega^{-n} {\tt t} ) (1 - \eta \omega^n)}  = 1$, i.e.
$$
({\tt t}^N-1)(\xi^N - \eta^N)=( \xi^N -1 ) ({\tt t}^N - \eta^N )- (\xi^N - {\tt t}^N ) (1 - \eta^N )= 0. 
$$
Hence ${\tt C}_{\xi, \eta}$ is singular if and only if $\xi^N = \eta^N$, in which case the kernel of ${\tt C}_{\xi, \omega^{-k} \xi}$ is one-dimensional space generated by the cyclic vector $v \in \CZ^N $ defined in (\req(v)). By (\req(ADCg)), for $\eta= \omega^{-k} \xi$ we have
$$
\begin{array}{lll}
{\tt C}_{\xi, \eta} (\omega^{-1} {\tt t}) X^{-1} {\tt A}_{\eta}({\tt t})v({\tt t}) &= 
{\tt C}_{\xi, \eta} (\omega {\tt t}) {\tt D}_{\xi}({\tt t})v({\tt t}) &=  0 , \\
v^*({\tt t}) {\tt A}_{\eta} ({\tt t})X^{-1}{\tt C}_{\xi, \eta}(\omega {\tt t}) &= v^*({\tt t}) {\tt D}_{\xi} ({\tt t}) {\tt C}_{\xi, \eta}(\omega^{-1} {\tt t}) &= 0.
\end{array}
$$
As $v (\omega^{-1} {\tt t})$ is characterized as a basis element of the kernel of ${\tt C}_{\xi, \eta} (\omega^{-1} {\tt t})$, (the same for $v (\omega {\tt t})$ as a basis of kernel of ${\tt C}_{\xi, \eta} (\omega {\tt t})$), there exist scalar functions $\lambda ({\tt t}), \lambda'({\tt t})$ so that the equalities hold:
$$
{\tt A}_{\eta}({\tt t})v({\tt t}) = \lambda ({\tt t}) X v (\omega^{-1} {\tt t}) , \ ~ \ {\tt D}_{\xi}({\tt t})v({\tt t}) = \lambda' ({\tt t}) v (\omega {\tt t}).
$$
Using (\req(v)) and the expression of ${\tt A}_{\eta}, {\tt D}_{\xi}$, one finds $\lambda({\tt t})= \frac{(\omega -  {\tt t} ) ( \xi \omega^{-k}   -  \omega^{-1}{\tt t} )  v({\tt t})_0}{( \xi - \omega^{-1} {\tt t} )v(\omega^{-1} {\tt t})_0}$, $\lambda'({\tt t}) = \frac{(1-{\tt t} )(\xi - {\tt t} ) v({\tt t})_0}{( \xi \omega^{-k} -{\tt t})v (\omega {\tt t})_0}$, then follows (\req(ADv)). By a similar argument, the one-dimensional cokernel of ${\tt C}_{\xi', \omega^{-k} \xi'}$ is generated by the vector $v^* \in \CZ^{N *} $ in (\req(v*)) satisfying the relation   
(\req(v*AD)).
\vspace{.1in} \par \noindent 
{\bf Remark}. When the 0th components of vectors $v, v^*$ in (\req(v)) (\req(v*)) with $k=0, \ldots, N-1$ are in the form 
\bea(ll)
v({\tt t})_0  &= c(\xi) \mu({\tt t}, \xi)^{-(N-k+1)} \prod_{l =1}^{N-k} ( \xi^{-1} {\tt t} - \omega^l )  , \\ 
v^*({\tt t})^0 & = c^*(\xi') \mu^* ({\tt t}, \xi')^{N-k+1} \prod_{l=0}^{N-k} (1- \xi'^{-1} {\tt t} \omega^{-l} )^{-1} ,   
\elea(v0)
with the functions $\mu({\tt t}, \xi), \mu^* ({\tt t}, \xi')  $ satisfying $\mu( \omega {\tt t}, \xi ) = \omega \mu({\tt t}, \xi)$, $\mu^* ({\tt t}, \omega \xi') = \omega \mu^* ({\tt t}, \xi') $, 
the relations (\req(ADv)), (\req(v*AD)) become
\bea(l)
{\tt A}_{\eta}({\tt t})v({\tt t}) = (1 -  \omega^{-1} {\tt t} ) X v (\omega^{-1} {\tt t}) , \ ~ \ {\tt D}_{\xi}({\tt t})v({\tt t}) = \omega (1-{\tt t} )  v (\omega {\tt t}) ; \\
v^*({\tt t}) {\tt A}_{\eta'} ({\tt t})  = (1-{\tt t}) v^*(\omega {\tt t}) X , \ ~ \
v^*({\tt t}) {\tt D}_{\xi'} ({\tt t}) = (\omega -{\tt t})  v^*(\omega^{-1} {\tt t}) .
\elea(v*ADv)
For example, the condition (\req(v0)) holds for $v({\tt t})_0  =   \xi {\tt t}^{-1} \prod_{l=1}^{N-k} ( \xi {\tt t}^{-1} - \omega^{-l} )$, $v^*({\tt t})^0 = \prod_{l =0}^{N-k} (1-\xi' {\tt t}^{-1} \omega^l )^{-1}$.  
\par \vspace{.2in} 
We shall consider only the cyclic vectors $v, v^*$ in (\req(v)), (\req(v*)) which satisfy the condition (\req(v0)). These vectors depend on the parameter $\xi, \xi'$ and $k \in \ZZ_N$, and will be denoted by $v({\tt t}) = v({\tt t}; \xi, k) \in \CZ^N$ and $v^*({\tt t}) = v^*({\tt t}; \xi', k) \in \CZ^{N *}$. By (\req(v)), $X^i v({\tt t}; \xi, k)$ is proportional to  $v({\tt t}; \omega^{-i}\xi, k)$ with the 0th component $(X^i v({\tt t}; \xi, k))_0$ satisfying the relation (\req(v0)) for $\omega^{-i}\xi$. We now determine the form of $c(\xi)$ in (\req(v0)) so that the equality holds:
$$
X^i v({\tt t}; \xi, k)= v({\tt t}; \omega^{-i}\xi, k) , \ i \in \ZZ_N . 
$$
It suffices to consider the case $i=-1$, which by (\req(v)), is equivalent to 
$$
c (\omega \xi) = \frac{( 1 -  \xi  ) }{(1 -  \omega^{1-k}\xi ) } c (\xi)  , ~ ~ \mu({\tt t}, \omega \xi)= \omega^{-1}\mu({\tt t}, \xi). 
$$
Up to $\xi^N$-function multiples,  $c (\xi)$ is equal to $\prod_{l=0}^{N-k} (1- \omega^l \xi)^{-1}$. Hence we may assume 
\be
v({\tt t}; \xi, k)_0 = \mu^{-(N-k+1)} \frac{ 1 }{1- \xi} \prod_{l =1}^{N-k} \frac{ \xi^{-1} {\tt t} - \omega^l }{1- \omega^l \xi} .
\ele(v0k)
Here $\mu = \mu ({\tt t}, \xi)$ is a variable algebraically depending on $({\tt t}, \xi)$ so that  $\mu({\tt t}, \omega^{-1} \xi)= \mu(\omega {\tt t},  \xi) = \omega \mu({\tt t}, \xi)$.
Similarly, the covector $v^*({\tt t}; \xi', k)$ with 
$$
v^*({\tt t}; \xi', k)  X^{-i} = v^*({\tt t}; \xi' \omega^{-i}, k)
$$
has the 0th component (\req(v0)) expressed by 
\be
v^*({\tt t}; \xi', k)^0  = \mu^* ({\tt t}, \xi')^{N-k+1} \frac{1}{1- \xi'^{-1} {\tt t}} \prod_{l=1}^{N-k} \frac{\xi' - \omega^{-l+1}}{1- \omega^{-l} \xi'^{-1} {\tt t}}, 
\ele(v*0k)
where $\mu^*({\tt t}, \omega^{-1} \xi')= \mu^*(\omega {\tt t},  \xi') = \omega \mu^*({\tt t}, \xi')$.
Furthermore, we shall require $\{ v({\tt t}; \xi, k) \}_{k} $, $\{ v^*({\tt t}; \xi', k) \}_k$ to be $N$-periodic for integers $k$, equivalently,  the variables $\mu= \mu({\tt t}, \xi)$, $\mu^*= \mu^* ({\tt t}, \xi')$ with the relations 
\be
(\xi^{-1} {\tt t})^N - 1= \mu^N(1- \xi^N), \ ~ \ ~ \mu^{* N}(\xi'^N - 1) =  1-  (\xi'^{-1} {\tt t})^N , 
\ele(muN)
i.e., $\{ v({\tt t}; \xi, k)_0 \}_{k \in \ZZ_N} , \{ v^*({\tt t}; \xi', k)^0\}_{k \in \ZZ_N} $ are cyclic $N$-vectors.

There exists a connection between the cyclic vectors $v({\tt t}; \xi, k), v^*({\tt t}; \xi', k)$ when the parameters $\xi, \xi'$ are related by $\xi \xi' = {\tt t}$, in which case one finds
$$
\frac{v_{-n+k}({\tt t}; \xi, k)}{v_{-n+k+1}({\tt t}; \xi, k)}   
=  \frac{(  1 - \omega^{n-k-1}\xi' ) ( \xi -\omega^{n-1}  ) }{(   1 -\omega^{n-1} \xi' ) (   \xi - \omega^{n-k}    ) }  = \frac{v^{* n}({\tt t}; \xi', k)}{v^{* n-1}({\tt t}; \xi', k)},
$$
equivalently, the covector $v^*({\tt t}; \xi', k)$ is proportionally related to the transport of $v({\tt t}; \xi, k)$ by
$$
v^*({\tt t}; \xi', k)^t \sim X^k J v({\tt t}; \xi, k), ~ ~ ~ ~ \xi \xi' = {\tt t},
$$
where $J$ is the $\CZ^N$-automorphism defined by $(Jv)_n = v_{-n}$ for $n \in \ZZ_N$. By (\req(v)) and (\req(v0k)), 
$$
\begin{array}{ll}
(X^kJ v({\tt t}; \xi, k))_0 &= v({\tt t}; \xi, k)_k \\
&= \frac{\mu({\tt t}, \xi)^{-(N-k+1)}}{(1 - \xi)} \prod_{l =1}^{N-k} \frac{ \xi^{-1} {\tt t} - \omega^l }{1- \omega^l \xi} \prod_{l=1}^k \frac{( \omega -\omega^l \xi  ) ({\tt t} -  \omega^{l-k} \xi)}{(1 -  \omega^{l-k}\xi ) ( {\tt t}- \omega^l \xi )} , 
\end{array} 
$$
then by (\req(v*0k)) (\req(muN)) and with the identification $\mu^* = (\omega \mu)^{-1}$ in (\req(muN)), one finds 
$$
\frac{1}{v({\tt t}; \xi, 1)_1} X^kJ v({\tt t}; \xi, k)  = ( \xi'- \omega  ) v^*({\tt t}; \xi', k)^t .
$$
Hereafter we will make the identification 
\be
\xi = {\tt x}, ~ ~ \xi' = {\tt y}, ~ ~ \xi \xi' ={\tt t}, ~ ~ \mu^* = (\omega \mu)^{-1},
\ele(xix')
the relation (\req(muN)) defines the algebraic surface:
\be
{\goth S} : ~ ~ {\tt y}^N - 1= \mu^N(1- {\tt x}^N), \ \ \sigma =({\tt x}, {\tt y}, \mu) \in \CZ^3 ~ ~ ~ {\rm with} \ \ {\tt t}:= {\tt x}{\tt y}.
\ele(S)
The element with ${\tt x}={\tt y} = \mu =1$ will be denoted by  
$${\sf 1}= (1, 1, 1).$$
 We shall use the Greek letters $\sigma, \sigma', \cdots$ to denote the surface elements in ${\goth S}$. For later use, we consider the following automorphisms of surface ${\goth S}$, 
\bea(ll)
U_1 : ({\tt x}, {\tt y}, \mu ) \mapsto (\omega {\tt x}, {\tt y}, \omega^{-1} \mu ) , & 
U_2 : ({\tt x}, {\tt y}, \mu ) \mapsto ( {\tt x}, \omega {\tt y}, \omega \mu ) , \\
V : ({\tt x}, {\tt y}, \mu ) \mapsto (\omega {\tt x}, \omega^{-1} {\tt y}, \omega^{-1}\mu ) , &
U : ({\tt x}, {\tt y}, \mu ) \mapsto (\omega {\tt x}, {\tt y}, \mu ) , \\
C : ({\tt x}, {\tt y}, \mu ) \mapsto ( {\tt y}, {\tt x}, \mu^{-1} ) .
\elea(AutS)
Now the vectors (\req(v)) (\req(v*)) with the 0th component, (\req(v0k)) and (\req(v*0k)) respectively, depend on the surface element $\sigma$ in (\req(S)) and $k \in \ZZ_N$, and we shall also write 
$$
v( \sigma ; k) = v ({\tt t}; \xi , k) \in \CZ^N , \ \ v^*( \sigma; k)= v^* ({\tt t}; \xi' , k) \in \CZ^{N*}. 
$$
Define  the functions of ${\goth S}$: 
\be
{\tt W}_\sigma (n) = \mu^{-n} \prod_{l =1}^{n} \frac{ {\tt y} - \omega^l }{1- \omega^l {\tt x}}, ~ ~ \overline{\tt W}_\sigma (n) = \mu^n \prod_{l=1}^n \frac{( \omega -\omega^l {\tt x}  ) }{( {\tt y}- \omega^l )} \ ~ ~ ~  ( \sigma \in {\goth S}) ,
\ele(WW-)
which satisfy the $N$-periodic condition for $n$, hence $n \in \ZZ_N$, and ${\tt W}_\sigma (0) = \overline{\tt W}_\sigma (0) = 1$. 
By (\req(v)) and (\req(v0k)), the vector $v ( \sigma ; k)$ is expressed  by 
\be
\mu (1 - {\tt x}) v_n ( \sigma ; k) = \overline{\tt W}_\sigma (n) {\tt W}_\sigma(n -k) .
\ele(vWW)
Also, by (\req(v*)) and (\req(v*0k)), one can express $v^{*}(\sigma, k)$ in terms of ${\tt W}_\sigma (n)$, $\overline{\tt W}_\sigma (n)$ in (\req(WW-)) by
\be
({\tt y} - \omega) v^{* n}(\sigma; k) = {\tt W}_\sigma(-n) \overline{\tt W}_\sigma (k-n) / \overline{\tt W}_\sigma (1) .
\ele(v*WW)
Note that $({\tt y} - \omega) \overline{\tt W}_\sigma (1) = \omega \mu (1 - {\tt x})$.

We now use cyclic vectors $v ( \sigma ; k), v^{* }(\sigma; k)$ to construct the $Q_R$, $Q_L$-operators. Set the diagonal matrix $\delta$ in (\req(Md)) by
\be
\delta_i = \omega^{-i} {\tt x} , ~ ~ i=0, \ldots, N-1 ,
\ele(delx)
and the $\CZ^N$-operator ${\tt S}_{i, j} (= {\tt S}_{i, j} (\sigma))$ for $i, j \in \ZZ_N$ by
\be
{\tt S}_{i, j} = {\tt v}_{i, j} \tau_{i, j}, ~ ~ ~ {\tt v}_{i,j} = X^i v (\sigma; j-i) (= v ( V^{-i} \sigma; j-i)) , \ \ \tau_{i, j} \in \CZ^{N *} .
\ele(Sij)
Here $v (\sigma; k)$ is the cyclic vector in (\req(vWW)), and $\tau_{i, j}$ is the parameter independent of $\sigma$. By the first two relations in (\req(v*ADv)), the $Q_R$-operator defined by (\req(QR)) satisfies the $TQ$-relation 
\be
\tau^{(2)} ( \omega^{-1} {\tt t} ) Q_R (\sigma) = (1 - \omega ^{-1} {\tt t} )^L X Q_R (U_2^{-1} \sigma) + \omega^L (1-{\tt t} )^L  Q_R (U_2 \sigma) ,
\ele(TQR)
where $U_2$ is the automorphism defined in (\req(QT)). 
We follows Baxter's mechanism in \cite{B72} to construct the companion of $Q_R$, the $Q_L$-operator, where the diagonal matrix $\delta$ in (\req(Md)) is set by
\be
\delta_i = \omega^{-i} {\tt y} , ~ ~ i=0, \ldots, N-1 .
\ele(dely)
Define the $\CZ^N$-operators $\widehat{\tt S}_{i, j} (= {\tt S}_{i, j} (\sigma ))$ for $i, j \in \ZZ_N$: 
\be
\widehat{\tt S}_{i, j}  = \widehat{\tau}_{i, j} \widehat{\tt v}_{i, j}, ~ ~ ~ 
\widehat{\tt v}_{i, j}= v^* ( \sigma; j-i) X^{-i} (= v^* ( V^i \sigma; j-i)) , \ ~ \widehat{\tau}_{i, j} \in \CZ^N , 
\ele(Shat)
and the $Q_L$-operator 
$$
Q_L (\sigma) = \sum_{i_\ell \in \ZZ_N} \otimes_{\ell=1}^L \widehat{\tt S}_{i_\ell, i_{\ell+1}} (\sigma), \ ~ \ ( L+1=1 ) . 
$$
By the last two relations in (\req(v*ADv)), the following $QT$-relation holds 
\be
Q_L (\sigma) \tau^{(2)} ( \omega^{-1} {\tt t} )  = (1 -  {\tt t} )^L Q_L ( U_1 \sigma) X  + (\omega -{\tt t} )^L  Q_L (U_1^{-1} \sigma)  .
\ele(QT) 
Note that the automorphism $U_1$ in (\req(QT)) is different from the $U_2$ in (\req(TQR)). 
As in \cite{Bax} (C28), we shall construct the $Q$-operator from $Q_R$ and $Q_L$ using the relation $
Q_L (\sigma) Q_R (\sigma') = Q_L (\sigma') Q_R (\sigma)$, which unfortunately fails for arbitrary $\sigma, \sigma' \in {\goth S}$. Nevertheless we shall look for the condition of $\sigma, \sigma'$ so that the commutative relation holds. Indeed, by the identification of variables,
\be
{\tt x} = \eta^{\frac{-1}{2}} x_q , \ ~ {\tt y} = \eta^{\frac{-1}{2}} y_q , \ ~ \mu = \mu_q , 
\ele(cod1)
where $\eta$ is defined in (\req(sup)), one may consider the curve (\req(xymu)) contained in the surface ${\goth S}$  (\req(S)) with equations invariant under automorphisms in (\req(AutS)):
\be
{\goth W}_{k'}: \ ~ ~ (1-k') {\tt x}^N = 1 - k' \mu^{-N} , ~ ~ (1-k') {\tt y}^N = 1 - k' \mu^N .
\ele(CPMc) 
In this way, the surface ${\goth S}$ is decomposed as the family of curves ${\goth W}_{k'}$ with the complex parameter $k'$. The base point ${\sf 1} \in {\goth S}$ is the superintegrable element $p$ in  (\req(sup)). Through the identification (\req(cod1)), (\req(CPMc)) and (\req(xymu)) are regarded  as two coordinate-systems of the same curve ${\goth W}_{k'}$, whose elements will be denoted by Roman letters $q, p, ...$ with coordinates $q=(x_q, y_q, \mu_q)$ in (\req(xymu)), and by letters $\sigma, {\sf p}, ...$  with the coordinates $\sigma= ({\tt x}, {\tt y}, \mu )$ in (\req(CPMc)). 
For an element $\sigma = q \in {\goth W}_{k'}$,  the Boltzmann weights (\req(CPW)) with $p$ in 
(\req(sup)) coincide with those in (\req(WW-)):
\be
W_{p, q} ( n ) = {\tt W}_\sigma (n) , ~ ~ ~ \overline{W}_{p, q} (n) = \overline{\tt W}_\sigma (n) .
\ele(SCPW)
For $\sigma=({\tt x}, {\tt y}, \mu) , \sigma'=({\tt x}', {\tt y}', \mu') \in {\goth W}_{k'}$, the Boltzmann weights in (\req(AutS)) become
\be
{\tt W}_{\sigma,\sigma'}(n)  = (\frac{\mu}{\mu'})^n \prod_{j=1}^n
\frac{{\tt y}'-\omega^j {\tt x}}{{\tt y}- \omega^j {\tt x}' }  , \ ~ \
\overline{\tt W}_{\sigma,\sigma'}(n) = ( \mu \mu')^n \prod_{j=1}^n \frac{\omega {\tt x} - \omega^j {\tt x}' }{ {\tt y}'- \omega^j {\tt y} } .
\ele(WS)
We now show
\be 
Q_L (\sigma) Q_R (\sigma') = Q_L (\sigma') Q_R (\sigma) ~ ~  \ ~ ~ \sigma, \sigma' \in {\goth W}_{k'} .
\ele(LQR)
Indeed we shall indicate the curve ${\goth W}_{k'}$ as the constraint condition for the above commutative property. Using the Baxter's method in \cite{B04, BBP, R06Q}, we  consider the product function: $f (\sigma, \sigma' | i, j ; k, l) = \widehat{\tt v}_{i, j}(\sigma) {\tt v}_{k, l} (\sigma')$, and look for an auxiliary function $p( \sigma, \sigma' | n)$ for $n \in \ZZ_N$ such that 
\be
p( \sigma, \sigma' | i - k ) f (\sigma, \sigma' | i, j ; k, l) p( \sigma, \sigma' | j- l )^{-1} = f (\sigma', \sigma | i, j ; k, l).
\ele(pfp)
By (\req(vWW)) and (\req(v*WW)), one finds 
$$
\begin{array}{ll}
&\omega \mu \mu' (1-{\tt x})(1- {\tt x}') f (\sigma, \sigma' | i, j ; k, l) \\
 =& \sum_{n \in \ZZ_N} \overline{\tt W}_\sigma (j-n) {\tt W}_\sigma(i-n) \overline{\tt W}_{\sigma'} (n-k) {\tt W}_{\sigma'} (n-l) \\
= & N \sum_{n \in \ZZ_N} V_{\sigma,\sigma'}(i, k ; n) V_{\sigma',\sigma}(-l, -j; n) ,
\end{array}
$$
where $V_{\sigma, \sigma'}(i, k ; n) := \frac{1}{N} \sum_{m \in\ZZ_N} \omega^{nm} {\tt W}_\sigma(i-m) \overline{\tt W}_{\sigma'} (m-k)$ ( \cite{BBP} (2.29)). In order to verify (\req(pfp)), it suffices to find another auxiliary function $\overline{p} ( \sigma, \sigma' | n)$ for $n \in \ZZ_N$ such that its Fourier transform $\overline{p}^{(f)} ( \sigma, \sigma' | n) (= \sum_{k=0}^{N-1} \omega^{nk} \overline{p}( \sigma, \sigma' | n))$ satisfies the relations for $n \in \ZZ_N$,
$$
p( \sigma, \sigma' | i - k ) V_{\sigma,\sigma'}(i, k ; n) = V_{\sigma',\sigma}(i, k ; n) \overline{p}^{(f)} ( \sigma, \sigma' | n) . 
$$ 
By summing up above relations for $n \in \ZZ_N$, the substitution, $i=j - j', k = j'' - j'$, in turn yields the following constraint of $p( \sigma, \sigma' | n ), \overline{p} ( \sigma, \sigma' | n)$'s: 
$$
{\tt W}_\sigma  (j- j') \overline{\tt W}_{\sigma'}(j' - j'') p( \sigma, \sigma' | j - j'' )  =  \sum_{n \in \ZZ_N} 
{\tt W}_{\sigma'} (j -n) \overline{\tt W}_{\sigma}(n- j'')    \overline{p} (\sigma, \sigma' |j'- n), 
$$
which is the star-triangle relation (\req(TArel)) with $p$ corresponding to ${\sf 1}$ and 
$$
p( \sigma, \sigma' | n )= {\tt W}_{\sigma \sigma'}(n), ~ ~   \overline{p} ( \sigma, \sigma' | n) = R_{p \sigma \sigma'}^{-1} \overline{\tt W}_{\sigma \sigma'}(n), \ ~ ~ \sigma, \sigma' \in {\goth W}_{k'} .
$$
By this, follows the relation (\req(pfp)) ( \cite{BBP} (2.30)), hence  (\req(LQR)). 
Note that $Q_R, Q_L$ take the $\infty$-value at the base element ${\sf 1}$. For convenience, we multiple the operators $Q_R, Q_L$ by normalized factors,
$$
\widetilde{Q}_R (\sigma ) = \mu^L (1-{\tt x})^L Q_R (\sigma ) , ~ ~ \widetilde{Q}_L (\sigma ) = \omega^L \mu^L (1-{\tt x})^L Q_L (\sigma ) ,
$$ 
so that the commutative relation (\req(LQR)) still holds:
\be
\widetilde{Q}_L (\sigma) \widetilde{Q}_R (\sigma') = \widetilde{Q}_L (\sigma') \widetilde{Q}_R (\sigma), \ ~ ~  \sigma, \sigma'\in {\goth W}_{k'}. 
\ele(LQR*)
Define the $Q$-operator
\be
Q(\sigma) = \widetilde{Q}_R (\sigma) \widetilde{Q}_R ({\sf 1})^{-1} = \widetilde{Q}_L ({\sf 1})^{-1} \widetilde{Q}_L (\sigma), \ \sigma \in {\goth W}_{k'} ,
\ele(Qdef)
when both $\widetilde{Q}_R ({\sf 1}), \widetilde{Q}_L ({\sf 1})$ are non-singular. Note that the $Q$-operator is independent of the choice of parameters $\tau_{i, j}, \widehat{\tau}_{i, j}$ in (\req(Sij)), (\req(Shat)) regardless of $\widetilde{Q}_R, \widetilde{Q}_L$ depending on them. 
By (\req(LQR*)) and (\req(TQR)), one finds 
$[Q(\sigma), Q(\sigma')]= [ \tau^{(2)}({\tt t}), Q(\sigma')] = 0$, and the $TQ$-relation
\be
\tau^{(2)} ( \omega^{-1} {\tt t} ) Q(\sigma) = (1 - \omega^{-1} {\tt t} )^L \omega^L X Q (U_2^{-1} \sigma) +  (1-{\tt t} )^L  Q (U_2 \sigma) .
\ele(TQ)
We now specify convenient parameters  $\tau_{i, j}, \widehat{\tau}_{i, j}$ for the explicit expression of the above $Q$-operator. Set $\tau_{i, j} = \langle j |$, $ \widehat{\tau}_{i, j}= | j \rangle$ in (\req(Sij)) (\req(Shat)), i.e.
\be
{\tt S}_{i, j} = X^i v ( \sigma; k) \langle j |, ~ ~ ~ \widehat{\tt S}_{i, j} = | j \rangle v^* ( \sigma; k) X^{-i} , ~ ~ (k= j-i \in \ZZ_N ).
\ele(Sdef)
By (\req(Sij)) and (\req(Sdef)), $
X{\tt S}_{i, j} (\sigma) X^{-1} = {\tt S}_{i+1, j+1} (\sigma )$, $ 
X^k {\tt S}_{i, j} (\sigma ) = 
{\tt S}_{i, j} (V^{-k} \sigma )$, which imply
\be
X Q_R (\sigma ) = Q_R (\sigma )X , ~ ~  X^k Q_R (\sigma ) = Q_R (V^{-k}\sigma ). 
\ele(XQR)
One can write a matrix expression of the $\widetilde{Q}_R$-operator using ${\tt W}_\sigma, \overline{\tt W}_\sigma$ in (\req(WW-)). Indeed by (\req(Sdef)), $
Q_R (\sigma ) |j'_1, \ldots, j'_L \rangle 
=  \otimes_{\ell=1}^L X^{j'_{\ell-1}} v ( \sigma ; j'_\ell-j'_{\ell-1})$  with $ L+ \ell = \ell$. The relation (\req(vWW)) yields 
$$
\langle j_1, \ldots, j_L| \widetilde{Q}_R (\sigma ) |j'_1, \ldots, j'_L \rangle  = \prod_{\ell=1}^L  {\tt W}_\sigma (j_\ell-j'_\ell) \overline{\tt W}_\sigma(j_{\ell+1}-j'_\ell) .
$$  
Similarly, the relations,  $
X \widehat{\tt S}_{i, j}(\sigma) X^{-1} = \widehat{\tt S}_{i+1, j+1}(\sigma)$ and $\widehat{\tt S}_{i, j}(\sigma) X = \widehat{\tt S}_{i, j}(V^{-1} \sigma )$, yield $X Q_L (\sigma ) = Q_L (\sigma )X $, $Q_L (\sigma) X^k = Q_L (V^{-k}\sigma )$. Using (\req(v*WW)), one finds 
$$
\langle j_1, \ldots, j_L|  \widetilde{Q}_L (\sigma) |j'_1, \ldots, j'_L \rangle = 
\prod_{\ell=1}^L 
\overline{\tt W}_\sigma (j_{\ell}-j'_\ell) {\tt W}_\sigma (j_\ell -j'_{\ell+1`}), 
$$
By (\req(SCPW)), $\widetilde{Q}_R$ and $\widetilde{Q}_L $ coincide respectively with the transfer matrices (\ref{Tpq}) (\ref{hTpq}) in CPM at the superintegrable point ${\sf 1}$:  
$$
\widetilde{Q}_R (\sigma) = T_{{\sf 1}, \sigma}, ~ ~ ~ \widetilde{Q}_L (\sigma) = \widehat{T}_{{\sf 1}, \sigma}.
$$  
When $\sigma = {\sf 1}$, ${\tt W}_{\sf 1}(n)=1, \overline{\tt W}_{\sf 1}(n) = \delta_{n, 0}$ for $n \in \ZZ_N$, hence $\widetilde{Q}_L ({\tt 1}) = {\rm Id}$, and the $Q$-operator (\req(Qdef)) is related to the CPM transfer matrix by $Q (\sigma) = \widehat{T}_{{\sf 1}, \sigma}$ for $\sigma \in {\goth W}_{k'}$. 
By (\req(XQR)), the $TQ$-relation (\req(TQ)) is equivalent to the following form (\cite{BBP} (4.20), \cite{R05o} (52)\footnote{The $N^L {\rm Q}_{cp}(q)$ in \cite{R05o} is equal to $\mu^L (1- {\tt x}^N)^L Q_R(\sigma) S_R$ here.})
$$
\tau^{(2)} ( \omega^{-1} {\tt t} ) Q_R (\sigma ) = (1 -  \omega^{-1} {\tt t} )^L Q_R (U^{-1} \sigma) + (1-{\tt t} )^L \omega^L X Q_R (U\sigma)
$$
where $U$ is the automorphism in (\req(AutS)).

\subsection{The $Q$-operator of the general superintegrable $\tau^{(2)}$-model  \label{ssec.QSUP}}
The construction of $Q_R, Q_L$ and $Q$-operators of $\tau^{(2)}$-model at the superintegrable point (\req(sup)) can be carried over to other superintegrable elements $p$ with $(x_p, y_p, \mu_p)= (\eta^{\frac{1}{2}}\omega^a,  \eta^{\frac{1}{2}} \omega^b, \omega^c )$, where the $L$-operators (\req(ttL)) with the parameter in (\req(cpmL)) for all $\eta$ are  gauge equivalent to 
\be
{\tt L} ( {\tt t} ) =  \left( \begin{array}{cc}
        1 - {\tt t} \omega^{2c} X  &   ( 1 -  \omega^{1+{\rm m}+2c} X)Z \\
        - {\tt t} (   1 - \omega^{{\rm m}+2c} X )Z^{-1} & - {\tt t} + \omega^{1+2{\rm m+2c}} X 
\end{array} \right) 
\ele(SUL) 
with ${\tt t}= t \eta^{-1} \omega^{-2b},  {\rm m} = a-b$. 
The change of coordinates\footnote{For the other type of superintegrable elements, $p=(x_p, y_p, \mu_p)= (\eta^{\frac{-1}{2}}\omega^a,  \eta^{\frac{-1}{2}} \omega^b, (-1)^{\frac{1}{N}}\omega^c )$, instead of (\req(codp)) we consider the change of coordinates: $({\tt x}, {\tt y}, \mu) = ( \eta^{\frac{1}{2}}\omega^{-b} x_q, \eta^{\frac{1}{2}}\omega^{-b} y_q,  (-1)^{\frac{-1}{N}} \mu_q)$. In this way, the curve ${\goth W}_{k'}$ in (\req(xymu)) is identified with ${\goth W}_{-k'}$ in (\req(CPMc)) with $p$ corresponding to ${\sf p}$ in (\req(supp)). } 
\be
{\tt x} = \eta^{\frac{-1}{2}} \omega^{-b} x_q , \ ~ {\tt y} = \eta^{\frac{-1}{2}} \omega^{-b} y_q , \ ~ \mu = \mu_q , 
\ele(codp)
provides two coordinate systems, (\req(CPMc)) and (\req(xymu)), for the curve ${\goth W}_{k'}$ in the surface ${\goth S}$ with the superintegrable element
\be
{\sf p}: ({\tt x}, {\tt y}, \mu) = (\omega^{\rm m}, 1, \omega^c) ~ ~ \Longleftrightarrow ~ ~ p: (x_p, y_p, \mu_p)= (\eta^{\frac{1}{2}}\omega^a,  \eta^{\frac{1}{2}} \omega^b, \omega^c ), 
\ele(supp)
where ${\rm m}:= a-b $.
We now consider the operator ${\tt A}_{\eta}, {\tt C}_{\xi, \eta}, {\tt D}_{\xi}$~$(\xi, \eta \in \CZ)$ in $(\req(gauL))$ associated to a $L$-operator (\req(SUL)). The Lemma \ref{lem:Ckck} is still valid by replacing  (\req(v)) (\req(v*)) respectively  by 
$$
\begin{array}{l}
\frac{v_n}{v_{n-1}}= \omega^{2c} \frac{( \omega^{{\rm m}+1}  -\omega^{n} \xi ) ({\tt t} - \omega^{{\rm m}+n-k} \xi)}{ (1 -  \omega^{n-k} \xi ) (  {\tt t} - \omega^{n} \xi  ) } , \\ 
\frac{v^{* n}}{v^{* n-1}} = \omega^{-2c}\frac{(1 -  \omega^{n-k-1} \xi') (  {\tt t} - \omega^{n-1} \xi'  ) }{ (\omega^{\rm m} - \omega^{n-1} \xi'  ) ({\tt t} - \omega^{{\rm m}+n-k} \xi')}, 
\end{array}
$$
and (\req(ADv)), (\req(v*AD)) by 
$$
\begin{array}{ll}
{\tt A}_{\omega^{-k} \xi }({\tt t})v({\tt t}) &= (1  - \omega^{-{\rm m}-1} {\tt t}) \frac{\omega^{2c+{\rm m}+1} ( {\tt t}- \omega^{{\rm m}-k+1} \xi )  v_0(t)  }{({\tt t}- \omega \xi   ) v (\omega^{-1} {\tt t})_0 }   X v (\omega^{-1} {\tt t}) , 
\\ 
{\tt D}_{\xi}({\tt t})v({\tt t}) &=  (  \omega^{\rm m}  -{\tt t} ) \frac {( {\tt t}-\xi    ) v({\tt t})_0 }{( {\tt t}- \omega^{{\rm m}-k} \xi  )v (\omega {\tt t})_0 }  v (\omega {\tt t}) ; \\
v^*({\tt t}) {\tt A}_{\omega^{-k} \xi'} ({\tt t}) &=
(1 -  \omega^{-{\rm m}} {\tt t} )\frac{ \omega^{2c+{\rm m}} (  \omega^{{\rm m}-k} \xi'-{\tt t} )v^{*0}({\tt t}) }{ (  \omega^{-1} \xi'-{\tt t}  )v^*(\omega t)^0  } v^*(\omega {\tt t}) X , \\
v^*({\tt t}) {\tt D}_{\xi'} ({\tt t})  &= ( \omega^{1+m}-{\tt t} )
 \frac{(   \xi'- {\tt t}   ) v^*({\tt t})^0 }{  ( \omega^{{\rm m}+1-k} \xi'-{\tt t} )v^*(\omega^{-1} t)^0} v^*(\omega^{-1} {\tt t}).
\end{array}
$$
With $\xi, \xi'$ identified with ${\tt x}, {\tt y}$ as in (\req(xix')), the cyclic vectors, $v( \sigma ; k)$ in (\req(vWW)) and $v^{*}(\sigma; k)$ in (\req(v*WW)), are given by the following general form
\bea(ll)
\mu^{1+ {\rm m}+ 2c} (\omega^{\rm m} - {\tt x}) v_n ( \sigma ; k) &= \overline{\tt W}_{{\sf p} \sigma} (n) {\tt W}_{{\sf p}, \sigma} (n -k)/{\tt W}_{{\sf p}, \sigma} (-{\rm m}), \\ 
\mu^{{\rm m}+ 2c} ({\tt y} - \omega^{1+{\rm m}}) v^{* n}(\sigma; k) &= {\tt W}_{{\sf p} \sigma} (-n) \overline{\tt W}_{{\sf p} \sigma}  (k-n)/ \overline{\tt W}_{{\sf p} \sigma}  (1+{\rm m}), 
\elea(vv*Gp)
for $\sigma$ in the surface ${\goth S}$ (\req(S)), where ${\tt W}_{\sigma \sigma'}, \overline{\tt W}_{\sigma \sigma'}$ are defined in (\req(WS)). Note that the following equality holds among the factors in above vectors:
$$
\mu^{{\rm m}+ 2c} ({\tt y} - \omega^{1+{\rm m}})\overline{\tt W}_{{\sf p} \sigma}  (1+{\rm m}) = \omega^{c(2 {\rm m}+1)+ {\rm m}({\rm m}+1)+1} \mu^{1+ {\rm m}+ 2c} (\omega^{\rm m} - {\tt x}) {\tt W}_{{\sf p}, \sigma} (-{\rm m}). 
$$
With the same argument in the previous subsection, we construct $Q_R, Q_L$-operators using the ${\tt S}, \widehat{\tt S}$-matrices in (\req(Sij)), (\req(Shat)), then identify the $Q$-operator with the CPM transfer matrix at the superintegrable point $p$ in (\req(supp)). We summarize the conclusion as follows.
\begin{thm}\label{thm:CPQp} 
Let $\tau^{(2)}({\tt t})$ be the matrix $(\req(tau2))$ associated to $L$-operator $(\req(SUL))$ at the superintegrable element ${\sf p}$ in $(\req(supp))$, and $T_{\sf p}, \widehat{T}_{\sf p}$ be the CPM transfer matrices in $(\ref{Tpq}), (\ref{hTpq})$ (through the identification $W_{p, q}= {\tt W}_{{\sf p} \sigma}$,$\overline{W}_{p, q}= \overline{\tt W}_{{\sf p} \sigma}$). Then the $Q_R, Q_L$-operators for the $\tau^{(2)}$-matrices are given by 
$$
Q_R (\sigma ) = \frac{T_{\sf p}(\sigma)}{ \mu^{1+ {\rm m}+ 2c} (\omega^{\rm m} - {\tt x}) {\tt W}_{{\sf p}, \sigma} (-{\rm m})} , ~ ~ Q_L(\sigma )= \frac{\widehat{T}_{\sf p} (\sigma)}{\mu^{{\rm m}+ 2c} ({\tt y} - \omega^{1+{\rm m}}) \overline{\tt W}_{{\sf p} \sigma}  (1+{\rm m})} 
$$
for $\sigma \in {\goth S}$, which satisfy the $TQ$-relation:
\bea(ll)
\tau^{(2)} ( \omega^{-1} {\tt t} ) Q_R (\sigma) = & (1  - \omega^{-{\rm m}-1} {\tt t})^L X Q_R (U_2^{-1} \sigma) + \omega^{(2c+{\rm m}+1)L} (  \omega^{\rm m}  -{\tt t} )^L Q_R (U_2 \sigma) , \\
Q_L (\sigma) \tau^{(2)} ( \omega^{-1} {\tt t} )  =&
(1 -  \omega^{-{\rm m}} {\tt t} )^L Q_L ( U_1 \sigma) X  + \omega^{(2c+{\rm m})L} ( \omega^{1+m}-{\tt t} )^L  Q_L (U_1^{-1} \sigma),  
\elea(TQpp)
where $U_1, U_2$ are automorphisms in $(\req(AutS))$. The relation $(\req(LQR))$, 
$Q_L (\sigma) Q_R (\sigma') = Q_L (\sigma') Q_R (\sigma)$, holds when $\sigma, \sigma'$ are in a CPM curve ${\goth W}_{k'}$, and the $Q$-operator defined by the normalized $Q_R, Q_L$-operators in $(\req(Qdef))$ is equal to the CPM transfer matrix at the superintegrable point ${\sf p}$: $
Q(\sigma) = \widehat{T}_{{\sf p}, \sigma} = T_{{\sf p}, \sigma} S_R $ for $\sigma \in {\goth W}_{k'}$.
\end{thm}
 \par \vspace{.2in}
It is known that the CPM possesses the Onsager-algebra symmetry for the superintegrable point ${\sf 1}$ \cite{R05o}; the same is also true for the superintegrable CPM  at an arbitrary superintegrable point in the above theorem. Indeed, 
as $\sigma$ tends to ${\sf p}$ in ${\goth W}_{k'}$ by setting ${\tt x}= \omega^{\rm m} (1- 2k' \epsilon + O(\epsilon^2))$ with small $\epsilon $, to the first order, one has ${\tt y}= 1 + 2k' \epsilon , \mu = 1+ 2(k'-1) \epsilon$, and  $\widehat{T}_{\sf p}$-expression for the $Q$-operator  near ${\sf p}$ ( \cite{AMP} (1.11)-(1.17) ):
$$
\widehat{T}_{\sf p} (\sigma) = {\bf 1} \{ 1 + (-1)^{\rm m}   (N-1-2{\rm m}) L \epsilon \} + \epsilon H + O(\epsilon^2) 
$$
where $H$ is the Hamiltonian expressed by
$$
H = k' H_0 + H_1 = (-1)^{{\rm m}+1} \bigg(k'\sum_{n=1}^{N-1}    \frac{2\omega^{n({\rm m}+2c)}}{1-\omega^{-n}} \sum_{\ell =1}^L X_\ell^n + \sum_{n=1}^{N-1}   \frac{2\omega^{n  {\rm m} }}{1-\omega^{-n}} \sum_{\ell =1}^L Z^n_\ell Z^{-n}_{\ell +1}  \bigg).   
$$
The above Hamiltonian with ${\rm m}= c =0$ was first found in \cite{GR}, where the operators $H_0, H_1$ were  shown to satisfy the Dolan-Grady relation, hence give rise to  a representation of Onsager algebra.  Indeed, the same argument also applies to the above Hamiltonian $H$ for two arbitrary integers  ${\rm m}, c$, hence the same conclusion holds for the Onsager-algebra representation using the Dolan-Grady pair, $H_0$ and $H_1$. Therefore one obtains the Onsager-algebra symmetry of the superintegrable $\tau^{(2)}$-model (\req(SUL)) by the same arguments as the case ${\rm m}= c =0$ in \cite{R05o}. Then the $TQ$-relation (\req(TQpp)) yields the following Bethe equation for the $\tau^{2}$-model:
\be
 \frac{(1- \omega^{-{\rm m}}t_i )^L}{(1- \omega^{-{\rm m}-1}t_i )^L}= - \omega^{-(2c+ 2 {\rm m} +1)L+ Q } \frac{F( \omega^{-1} t_i ) }{F( \omega t_i )} 
\ele(Bethe)
for $i=1, \ldots, J$, where $F({\tt t}):= \prod_{j=1}^J (1- t_j^{-1} {\tt t})$ .

\section{The $Q$-operator of XXZ Chain at a Root of Unity \label{sec:QXXZ}}
\setcounter{equation}{0}
In this section, we employ the theory of cyclic representations of quantum algebra $U_{\sf q} (sl_2)$ at $N$th root-of-unity ${\sf q}$ to study the transfer matrix of XXZ chain, and establish an equivalent relationship between these models and an one-parameter family of generalized $\tau^{(2)}$-model. For the case of cyclic representations with $N$th root-of-unity representation-parameter for odd $N$, we construct the $Q$-operator for the corresponding $\tau^{(2)}$-model then identify it with the (inhomogeneous) CPM transfer matrices for two vertical superintegrable rapidities. In particular, for the special cyclic representation describing the spin-$\frac{N-1}{2}$ representation of $U_{\sf q}(sl_2)$, the result provides the identical theory between  the root-of-unity XXZ chain of spin-$\frac{N-1}{2}$ and the homogeneous superintegrable CPM at a specific point.

\subsection{XXZ chain, cyclic representations of $U_{\sf q} (sl_2)$ and generalized $\tau^{(2)}$-model \label{ssec.cyXXZ}}
The quantum algebra $U_{\sf q} (sl_2)$, i.e. the associated $\CZ$-algebra generated by $K^\frac{\pm 1}{2}, e^{\pm}$ with the relations $K^{\frac{1}{2}} K^{\frac{-1}{2}}= K^{\frac{-1}{2}}K^{\frac{1}{2}}=1$ and 
\be
 K^{\frac{1}{2}} e^{\pm } K^{\frac{-1}{2}} = {\sf q}^{\pm 1} e^{\pm} , ~ ~ ~ [e^+ , e^- ] = \frac{K-K^{-1}}{{\sf q}- {\sf q}^{-1}} ,
\ele(Uqsl)
arises from the theory of six-vertex model as the YB solution for the $R$-matrix 
$$
R_{\rm 6v} (s) = \left( \begin{array}{cccc}
        s^{-1} {\sf q} - s {\sf q}^{-1}  & 0 & 0 & 0 \\
        0 &s^{-1} - s & {\sf q}-{\sf q}^{-1} &  0 \\ 
        0 & {\sf q}-{\sf q}^{-1} &s^{-1} - s & 0 \\
     0 & 0 &0 & s^{-1} {\sf q} - s {\sf q}^{-1} 
\end{array} \right)
$$   
\cite{Fad, KBI, KRS, KS}.
The solution, called the $L$-operator, is the matrix with entries in $U_{\sf q}(sl_2)$, 
\be
{\cal L} (s)  =  \left( \begin{array}{cc}
        {\cal A} (s)  & {\cal B} (s) \\
        {\cal C} (s) & {\cal D} (s)
\end{array} \right) :=  \left( \begin{array}{cc}
        s K^{\frac{-1}{2}} - s^{-1} K^{\frac{1}{2}}  & ({\sf q}-{\sf q}^{-1}) e^-  \\
        ({\sf q}-{\sf q}^{-1}) e^+ & s K^{\frac{1}{2}} - s^{-1} K^{\frac{-1}{2}}
\end{array} \right) 
\ele(6VL)
for $s \in \CZ$, which satisfies the YB equation 
\be
R_{\rm 6v}(s/s') ({\cal L}(s) \bigotimes_{aux}1) ( 1
\bigotimes_{aux} {\cal L}(s')) = (1
\bigotimes_{aux} {\cal L}(s'))( {\cal L}(s)
\bigotimes_{aux} 1) R_{\rm 6v}(s/s').
\ele(6YB)
Indeed, the YB constraint (\req(6YB)) for ${\cal L}$ in the form (\req(6VL)) is equivalent to the  relation (\req(Uqsl)) for the algebra $U_{\sf q} (sl_2)$. Since (\req(6YB)) is still valid when changing the variable $s$ by $\lambda s$ using a non-zero complex $\lambda$, the matrix $L(s) = r ({\cal L} (\lambda s))$ for a representation $r : U_{\sf q} (sl_2) \longrightarrow {\rm End} (\CZ^d)$  becomes a $L$-operator with $\CZ^2$-auxiliary and $\CZ^d$-quantum space satisfying the YB relation (\req(6YB)). In particular, for $\lambda = {\sf q}^{\frac{d-2}{2}}$ and $r$ the spin-$\frac{d-1}{2}$ (highest-weight) representation of $\CZ^d = \oplus_{k=0}^{d-1} \CZ {\bf e}^k $:
\be
K^{\frac{1}{2}} ({\bf e}^k) = {\sf q}^{\frac{d-1-2k}{2}} {\bf e}^k , \ \ e^+ ( {\bf e}^k ) = [ k  ] {\bf e}^{k-1} , \ \ e^-( {\bf e}^k ) = [ d-1-k ] {\bf e}^{k+1}  ,
\ele(drep)
where $[n]= \frac{{\sf q}^n - {\sf q}^{-n}}{{\sf q}-{\sf q}^{-1}}$ and $ e^+ ( {\bf e}^{0} ) = e^- ( {\bf e}^{d-1} )= 0$, one obtains the well-known $L$-operator of XXZ chain of spin-$\frac{d-1}{2}$ (see, e.g. \cite{KiR, R06Q, R06F, R07} and references therein). Using the local $L$-operator (\req(6VL)), one constructs the monodromy matrix
 $\bigotimes_{\ell=1}^L {\cal L}_\ell (s)$ with entries in $(\stackrel{L}{\bigotimes} U_{\sf q}(sl_2))(s)$,
$$
\bigotimes_{\ell=1}^L {\cal L}_\ell (s)  =  \left( \begin{array}{cc}
        {\cal A}_L (s)  & {\cal B}_L (s) \\
        {\cal C}_L (s) & {\cal D}_L (s)
\end{array} \right) 
$$
again satisfying (\req(6YB)). Denote the leading and lowest terms of entries of the above  monodromy matrix,  
$$
\begin{array}{ll}
{\sf A}_\pm = \lim_{s^{\pm} \rightarrow \infty} (\pm s)^{\mp L}{\cal A}_L(s) , & {\sf B}_\pm = \lim_{s^{\pm} \rightarrow \infty} (\pm s)^{\mp (L-1)} \frac{{\cal B}_L(s)}{q-q^{-1}} ,  \\ 
{\sf C}_\pm = \lim_{s^{\pm} \rightarrow \infty} (\pm s )^{\mp (L-1)}\frac{{\cal C}_L(s)}{q-q^{-1}}, & {\sf D}_\pm = \lim_{s^{\pm} \rightarrow \infty} (\pm s)^{\mp L}{\cal D}_L(s),
\end{array}  
$$ by
$$
q^{ \pm S^z}= {\sf A}_\mp = {\sf D}_\pm , \ ~ \ T^- = {\sf B}_+ , \  S^- = {\sf B}_- , \ ~ \ S^+ = {\sf C}_+ , \ T^+ = {\sf C}_- .
$$
The above elements give rise to the infinite-quantum-algebra $U_q(\widehat{sl}_2)$ with the identification:
$$
k_0^{-1} = k_1 = q^{2 S^z}, ~ e_0 = T^-   ,  f_0 = T^+ , ~ 
e_1 = S^+  , ~  f_1 = S^- . 
$$
When applying a $U_{\sf q}(sl_2)$-representation and scaling the variable $s$ by $\lambda s$, 
the traces,   
${\rm tr}_{\CZ^2} \bigotimes_{\ell=1}^L {\cal L}_\ell (s)$, form a commuting family for all $s$, which gives rise to the transfer matrix of XXZ chain. In particular, by
employing the spin-$\frac{d-1}{2}$ representation of $U_{\sf q}(sl_2)$ and setting $\lambda = {\sf q}^{\frac{d-2}{2}}$, one obtains the commuting $\stackrel{L}{\otimes} \CZ^d$-operator for
the spin-$\frac{d-1}{2}$  XXZ model (see, e.g. \cite{R06F} section 4.2).

In the root of unity case when ${\sf q}$ is a $N$th primitive root-of-unity ${\sf q}$, as in the study of root-of-unity spin-$\frac{1}{2}$ XXZ chain in \cite{DFM}, the normalized $N$th power of $S^\pm, T^\pm$, $S^{\pm (N)} = \frac{S^{\pm N}}{[N]!}$ , $T^{\pm (N)} = \frac{T^{\pm N}}{[N]!}$ ( $[N]!:= \prod_{i=1}^N [i]$), and $\frac{2S^z}{N}$ generate the $sl_2$-loop-algebra by 
$$
- H_0 = H_1 = \frac{2S^z}{N} ; \ E_0 = T^{- (N)}, \ E_1 = S^{+ (N)}, \ F_0 = T^{+ (N)},  \ F_1 = S^{- (N)}.
$$
Using the above relation, one can relate the finite algebra $U_{\sf q} (sl_2)$ to the affine loop algebra of $sl_2$. We now consider
the  cyclic representations $\rho_\varepsilon$ of $U_{\sf q} ( sl_2)$, labelled by an arbitrary complex parameter $\varepsilon$ (see, e.g. \cite{DJMM} and references therein), where the $\rho_\varepsilon$-states are $N$-cyclic vectors  in $\CZ^N$ with the $U_{\sf q} ( sl_2)$-action:
\be
K |n \rangle = {\sf q}^{-2n} |n \rangle , \ ~ ~  \ e^\pm  |n \rangle = \frac{ {\sf q}^{\varepsilon \pm n}- {\sf q}^{- \varepsilon \mp n}  }{{\sf q}-{\sf q}^{-1}} |n \mp 1 \rangle, ~ ~ ~ n \in \ZZ_N .
\ele(crep)
The matrix $L(s) = \rho_\varepsilon {\cal L}(s)$ is the $L$-operator for the transfer matrix of the XXZ chain with the cyclic representation $\rho_\varepsilon$,
\be
T (s) =  \rho_\varepsilon \bigg( {\rm tr}_{\CZ^2} \bigotimes_{\ell=1}^L {\cal L}_\ell \bigg) (s) .
\ele(TcXZ)
Note that $T(s)$ commutes with $K (:= \otimes_\ell K_\ell$, the product of local $K$'s).

We now consider the cases for odd $N= 2M+1$. First, we note that the spin-$\frac{N-1}{2}$ representation (\req(drep)) can be regarded as the cyclic representation $\rho_{\varepsilon = M}$ (i.e. ${\sf q}^{\varepsilon} = {\sf q}^{\frac{-1}{2}}$), where the basis elements ${\bf e}^k$ in (\req(drep)) are identified with $|-M+k \rangle$ in (\req(crep)) for $k=0, \ldots, N-1$. Now we choose the primitive $N$th root-of-unity ${\sf q}$ with ${\sf q}^{-2} = \omega$. One can use the Weyl operators $X, Z$ to present the cyclic representation (\req(crep)): $K = Z$ , $({\sf q}-{\sf q}^{-1}) e^{\pm}  = ( {\sf q}^{\varepsilon+1} Z^{\mp \frac{ 1}{2}} - {\sf q}^{-\varepsilon-1} Z^{\pm \frac{ 1}{2}})X^{\mp 1}$; so is the $L$-operator $L (s)  (= \rho_\varepsilon {\cal L}(s))$:
$$
L (s) 
= - s^{-1} Z^{\frac{1}{2}} \left( \begin{array}{cc}
      1-  s^2 Z^{-1}   & - s {\sf q}^{\varepsilon+1} ( 1  - {\sf q}^{-2\varepsilon-2} Z^{-1}) X  \\
      s {\sf q}^{-\varepsilon-1}   ( 1- {\sf q}^{2\varepsilon+2} Z^{-1}  )X^{-1} &- s^2  + Z^{-1}
\end{array} \right) .  
$$
Therefore $-s K^{\frac{-1}{2}}L (s)$ is gauge equivalent to 
$$
\left( \begin{array}{cc}
      1-  {\tt t} Z^{-1}   &  ( 1  - {\sf q}^{-2\varepsilon-2} Z^{-1}) X  \\
     - s^2  ( 1- {\sf q}^{2\varepsilon+2} Z^{-1}  )X^{-1} &- {\tt t}  + Z^{-1}
\end{array} \right) . 
$$
Since the pair of Weyl operators  $Z^{-1}, X$ can be converted to $X, Z$ through the change of $\CZ^N$-basis $| n \rangle' = \frac{1}{N} \sum_{j \in \ZZ_N} \omega^{-nj} |j \rangle$: 
\be
\bigg( {}^{Z^{-1}}_X |0 \rangle', \cdots, {}^{Z^{-1}}_X | N-1 \rangle' \bigg) = \bigg( |0 \rangle', \cdots, | N-1 \rangle' \bigg) {}^X_{Z} , 
\ele(basC)
the above matrix is isomorphic to the $L$-matrix (\req(ttL)) of $\tau^{(2)}$-model with  
$-\alpha = \gamma = - \kappa = 1,  \varrho = \varsigma^{-1}, -\beta = \varsigma \in \CZ$, i.e.,
$- s K^{\frac{-1}{2}} L (s)$ is equivalent to  
\be
{\tt L}({\tt t})= \left( \begin{array}{cc}
      1-  {\tt t} X   &  ( 1  - \varsigma^{-1} X) Z  \\
     - {\tt t}  ( 1- \varsigma X  )Z^{-1} &- {\tt t}  + X
\end{array} \right), ~ \varsigma \in \CZ  
\ele(Ktau)
where ${\tt t}= s^2$ and $\varsigma:= {\sf q}^{2\varepsilon+2}$.
Note the $L$-operators in (\req(Ktau)) differ from $\tau^{(2)}$-models of homogeneous CPM (\req(cpmL)) except only at $\varsigma = \omega^{\frac{-1}{2}} (=\omega^M)$, equivalently ${\sf q}^{\varepsilon}= {\sf q}^{\frac{-1}{2}}$, which corresponds to the spin-$\frac{N-1}{2}$ representation (\req(drep)) 
where $L ({\sf q}^{\frac{N-2}{2}} s)$ is the $L$-operator for the spin-$\frac{N-1}{2}$ XXZ chain in \cite{R06F}. Furthermore, (\req(Ktau)) with $\varsigma = \omega^M$ is the same as the superintegrable $\tau^{(2)}$-model (\req(SUL)) (with ${\rm m}=M, c=0$)  at ${\sf p}= (\omega^M, 1, 1)$,  whose $Q$-operator is the CPM transfer matrix at ${\sf p}$ by Theorem  \ref{thm:CPQp}. Hence we obtain the following result.
\begin{thm}\label{thm:crepT} 
Let $L(s) = \rho_\varepsilon {\cal L}(s)$ is the $L$-operator for the transfer matrix $T(s) ~ (\req(TcXZ))$ of XXZ chain associated to the cyclic representation $\rho_\varepsilon$ of $U_{\sf q}(sl_2)$ for the $N$th root-of-unity ${\sf q}= \omega^{\frac{-1}{2}}$ with odd $N=2M+1$. Then $- s K^{\frac{-1}{2}} L (s) $ is equivalent to the $L$-operator $(\req(Ktau))$ of the $\tau^{(2)}$-model parametrized by $\varsigma \in \CZ$. Under the conjugation of local-basis change in $(\req(basC))$, one has the identification
\be
- s^L K^{\frac{-1}{2}} T(s) \equiv \tau^{(2)}( \omega^{-1} {\tt t}) , \ \ {\tt t}= s^2 , ~ ~ \varsigma = {\sf q}^{2\varepsilon+2} .
\ele(Ttau)
In particular when ${\sf q}^{\varepsilon} = {\sf q}^{\frac{-1}{2}}$ (i.e. $\varsigma = \omega^{\frac{-1}{2}}$)  where $\rho_\varepsilon$ becomes the spin-$\frac{N-1}{2}$ representation of $U_{\sf q}(sl_2)$, the relation $(\req(Ttau))$ provides the identical theory between the XXZ chain of spin-$\frac{N-1}{2}$ with the anisotropic parameter ${\sf q}$ and superintegrable CPM at ${\sf p}= (\omega^M, 1, 1)$ so that the CPM transfer matrix $(\ref{hTpq})$ (or $(\ref{Tpq})$) serves as the $Q$-operator of the spin-$\frac{N-1}{2}$ XXZ chain.
\end{thm}
 \par \noindent 
{\bf Remark}. Through the identification of XXZ and CPM in the above theorem, 
the substitution ${\tt t} =  {\sf q}^{-2}s^2$ in (\req(Bethe)) with $({\rm m}, c)=(M, 0)$, i.e. the Bethe equation of $\tau^{(2)}$-model for ${\sf p}= (\omega^M, 1, 1)$, yields the following Bethe equation of the spin-$\frac{N-1}{2}$ XXZ chain (see, e.g. \cite{R06F} (4.22)):
$$
 \frac{a(s_i)^L}{a({\sf q}^{N-1}s_i )^L}= - {\sf q}^{2 Q -L}\prod_{j=1}^J \frac{( s_j^2 - {\sf q}^{-2} s_i^2 )}{ ( s_j^2- {\sf q}^2  s_i^2 ) }  
$$
for $i=1, \ldots, J$, where $a(s):= s {\sf q}^{\frac{-1}{2}} - s^{-1} {\sf q}^{\frac{1}{2}}$. The root-of-unity spin-$\frac{N-1}{2}$ XXZ chain is known to possess the $sl_2$-loop-algebra symmetry \cite{R06F}. The identical theory between XXZ and CPM in Theorem \ref{thm:crepT} implies that  
the Onsager-algebra symmetry found in the superintegrable CPM in \cite{R05o}, (or more precisely, in section \ref{ssec.QSUP} of this paper about the CPM at the superintegrable point $(\omega^M, 1, 1)$), can be extended to the $sl_2$-loop-algebra symmetry; equivalently to say, the root-of-unity spin-$\frac{N-1}{2}$ XXZ chain indeed inherits the Onsager-algebra symmetry from the $Q$-operator, compatible with the $sl_2$-loop-algebra symmetry as indicated by the general representation theory of these algebras in \cite{R91}. This finding provides a satisfactory answer to the question in section 4.3 of \cite{R06F}  about the symmetry structure of these two models.

\subsection{Inhomogeneous superintegrable CPM transfer matrix as the $Q$-operator of  a generalized $\tau^{(2)}$-model \label{ssec.IcpQ}}
In this subsection, we identify the CPM transfer matrix with two specific vertical superintegrable rapidities as the $Q$-operator of the $\tau^{(2)}$-model (\req(Ktau)) with $\varsigma^N=1$ for a positive integer $N$ (no oddness required). In particular when $N$ is odd, the result provides the relation between (inhomogeneous) superintegrable $N$-state CPM and the XXX chain for cyclic representations of $U_{\sf q} (sl_2)$ with ${\sf q}^N = {\sf q}^{\varepsilon N} = 1$ by Theorem \ref{thm:crepT}. As in section \ref{sec:Qtau}, we first determine the kernel of ${\tt C}_{\xi, \eta}$-operator in (\req(gauL)) associated to the $L$-operator (\req(Ktau)).
The entries of ${\tt C}_{\xi, \eta}$ are zeros except
$$
\langle  n | {\tt C}_{\xi, \eta}| n \rangle = (\xi - \omega^{-n} {\tt t} ) (1 - \eta \omega^n) , ~ \ ~  \langle  n | {\tt C}_{\xi, \eta}| n-1 \rangle = -( \xi - \omega^{-n+1}\varsigma ) ( {\tt t} - \eta \omega^{n-1} \varsigma^{-1}) , 
$$
hence the criterion of $\xi, \eta$ for the existence of a non-zero kernel vector is : $\xi^N = \eta^N$  when $\varsigma^N = 1$, and $(\xi^N , \eta^N) = (\varsigma^N, 1), (-1, - \varsigma^N)$ when $\varsigma^N \neq 1$. Hereafter, we shall consider only the case $\varsigma^N = 1$, 
(for odd $N$, corresponding to $T(s)$ in (\req(TcXZ)) with a cyclic representation $\rho_\varepsilon$ and $\varepsilon \in \frac{1}{2}\ZZ$). As in Lemma \ref{lem:Ckck}, one can show the following result.
\begin{lem}\label{lem:6VcC}
Let ${\tt C}_{\xi, \eta}$ be the operator in $(\req(gauL))$ associated to the $L$-operator $(\req(Ktau))$ with $\varsigma^N =1$. Then ${\tt C}_{\xi, \eta}$ is singular if and only if $\xi^N = \eta^N$. When $\eta = \omega^{-k} \xi $,  ${\tt C}_{\xi, \eta}$ has one-dimensional kernel with the cyclic-vector basis  $w = \sum_{n \in \ZZ_N} w_n |n \rangle \in \CZ^N$ determined by
\be
\frac{w_n}{w_{n-1}} = \frac{( \varsigma \omega  - \omega^n \xi  ) ({\tt t} -  \varsigma^{-1} \omega^{n-k -1} \xi )}{  (1 -  \omega^{n-k} \xi ) (  {\tt t} - \omega^n \xi  )}   \ ~ \ ( n \in \ZZ_N) ,
\ele(w)
satisfying the relations
\bea(l)
{\tt A}_{\eta}({\tt t})w({\tt t}) = ( \varsigma \omega    -  {\tt t}) \frac { ( \varsigma^{-1} \omega^{-k } \xi - {\tt t})w({\tt t})_0} {  (   \omega \xi - {\tt t}   )w (\omega^{-1} {\tt t})_0 }  X w (\omega^{-1} {\tt t}) , 
\\ {\tt D}_{\xi}({\tt t}) w({\tt t}) = ( \varsigma^{-1}\omega^{-1} - {\tt t})  \frac{ (\xi-{\tt t}) w({\tt t})_0 }{  (\varsigma^{-1} \omega^{-k -1} \xi - {\tt t})w (\omega {\tt t})_0 }  w (\omega {\tt t}) .
\elea(ADw6)
For $\eta' = \omega^{-k} \xi' $, the one-dimensional cokernel of ${\tt C}_{\xi', \eta'}$ is generated by $w^* = \sum_{n \in \ZZ_N} w^{* n} \langle n | \in \CZ^{N *}$ expressed by
\be
\frac{w^{* n}}{w^{* n-1}} = \frac{ (1 -  \omega^{n-k-1} \xi' )(  {\tt t} -  \omega^{n-1} \xi'  )}{ (  \varsigma  -  \omega^{n-1} \xi') ({\tt t} -  \varsigma^{-1} \omega^{n-k-1}  \xi' )}  \ ~ \ ( n \in \ZZ_N) , 
\ele(w*)
satisfying the relations
\bea(l)
w^*({\tt t}) {\tt A}_{\eta'} ({\tt t})  =   (  \varsigma  -  {\tt t}  ) \frac{  ( \varsigma^{-1} \omega^{-k-1}  \xi' - {\tt t}) w^{*0}({\tt t})} {  (    \omega^{-1} \xi' -{\tt t} ) w^{* 0}(\omega {\tt t})}  w^*(\omega {\tt t}) X , \ ~ \
\\
w^*({\tt t}) {\tt D}_{\xi'} ( {\tt t}) = (\varsigma^{-1} -{\tt t} )  \frac{(\xi' - {\tt t}) w^{*0}({\tt t})}{ (  \varsigma^{-1} \omega^{-k}  \xi' - {\tt t} )w^{*0}(\omega ^{-1} {\tt t}) }  w^* (\omega ^{-1} {\tt t}) .
\elea(w*AD6)
\end{lem}
 \par \vspace{.2in}
With the identification of the variables $\xi, \xi'$ with ${\tt x}, {\tt y}$ in (\req(xix')), we now describe the special cyclic vectors $w( \sigma; k) (= w({\tt t}; {\tt x},  k ))$ , $w^*( \sigma; k) (= w^*({\tt t}; {\tt y},  k ))$ in (\req(w)), (\req(w*)) respectively, for $\sigma$ in the surface ${\goth S}$ (\req(S)) so that (\req(ADw6)) and (\req(w*AD6)) will take a simple form. Consider the following elements in ${\goth S}$:
\bea(lll)
{\sf p}: ({\tt x}_{\sf p}, {\tt y}_{\sf p}, \mu_{\sf p})= (\varsigma^{-1} \omega^{-1} , 1, 1) , & {\sf p}' : ({\tt x}_{{\sf p}'}, {\tt y}_{{\sf p}'}, \mu_{{\sf p}'})= (\varsigma , 1, 1) , &
\varsigma= \omega^m  
\elea(p'p)
where $m \geq 0$ an non-negative integer.
Instead of (\req(WW-)), we use the functions ${\tt W}_{{\sf p}, \sigma}(n), \overline{\tt W}_{{\sf p},\sigma}(n)$, ${\tt W}_{{\sf p}', \sigma}(n), \overline{\tt W}_{{\sf p}',\sigma}(n)$ in (\req(WS)) for $\sigma = ({\tt x}, {\tt y}, \mu) \in {\goth S}, n \in \ZZ_N$.
Note that by the definition of ${\sf p}, {\sf p}'$ in (\req(p'p)), these functions are indeed defined on ${\goth S}$ (no constraint of the curve (\req(CPMc))). Define the cyclic vector $w( \sigma; k)$ in (\req(w)), and $w^* ( \sigma; k)$ in (\req(w*)) by
\bea(l)
\mu^{m+1} \bigg( \prod_{l=-m}^{m+1} (1- \omega^l {\tt x}) \bigg) w( \sigma; k)_n = 
\overline{\tt W}_{{\sf p}',\sigma}(n) {\tt W}_{{\sf p},\sigma}(n- k)/ {\tt W}_{{\sf p},\sigma}(m+1), ~    \\
\mu^{m} \bigg( \prod_{l = -m}^{m-1} (\omega^l{\tt y}-1)^{-1} \bigg)  w^{*} ( \sigma; k)^n = 
{\tt W}_{{\sf p}',\sigma }( -n ) \overline{\tt W}_{{\sf p},\sigma }(k -n )/ \overline{\tt W}_{{\sf p},\sigma }(-m) . 
\elea(ww*k)
(The left factor of the second formula in above is set to be $1$ when $m=0$.)
Then the relations (\req(ADw6)), (\req(w*AD6)) take the form 
\bea(ll)
{\tt A}_{\omega^{-k}{\tt x}}({\tt t})w( \sigma)  &= ( 1    - \varsigma^{-1} \omega^{-1} {\tt t})  X w( U^{-1}_2 \sigma)  , \\
{\tt D}_{\tt x} ({\tt t}) w( \sigma; k) &= ( 1 - \varsigma \omega {\tt t})  w( U_2 \sigma) , \\ 
w^* ( \sigma; k) {\tt A}_{\omega^k {\tt y} } ({\tt t})  &=  (  1  - \varsigma^{-1} {\tt t}  ) w^* ( U_1 \sigma) X , \\
w^* ( \sigma) {\tt D}_{\tt y}  ( {\tt t}) & =  ( 1 - \varsigma {\tt t} )  w^* ( U_1^{-1} \sigma) ,
\elea(AD6)
with the following relations for the cyclic vectors,
$$
X^i w( \sigma; k) = w( V^{-i} \sigma; k) , ~ ~ ~ \
w^{*} ( \sigma; k) X^{-i} = w^{*} ( V^i \sigma; k) ,  
$$
where $U_1, U_2, V$ are automorphisms in (\req(AutS)).

The $Q_R$, $Q_L$-operators for the $L$-operator (\req(Ktau)) are constructed as in the case of the superintegrable $\tau^{(2)}$-model in section \ref{ssec.Qsup1}.  We replace $v (\sigma; j-i)$, $v^* ( \sigma; j-i)$ in (\req(Sij)), (\req(Shat)) by $w (\sigma; j-i), w^* ( \sigma; j-i)$ in (\req(ww*k)) respectively, then obtain the ${\tt S}$, $\widehat{\tt S}$-matrices, hence the $Q_R$, $Q_L$-operators. The relation (\req(AD6)) in turn yields the following $TQ$-relation  
\bea(ll)
\tau^{(2)}(\omega^{-1} {\tt t} ) Q_R (\sigma) = &( 1    -  \varsigma^{-1} \omega^{-1} {\tt t})^L X Q_R (U_2^{-1} \sigma) + ( 1 - \varsigma \omega {\tt t})^L  Q_R (U_2 \sigma) ; \\
Q_L (\sigma) \tau^{(2)}(\omega^{-1} {\tt t} )  =&(  1  -  \varsigma^{-1} {\tt t}  )^L Q_L ( U_1 \sigma) X  + (1 - \varsigma {\tt t} )^L  Q_L (U_1^{-1} \sigma)  .
\elea(TQa)
Note that by (\req(ww*k)), the $Q_R, Q_L$-operators take the $\infty$- or zero-values at ${\sf p}, {\sf p}'$. We multiple the $Q_R, Q_L$-operators by the factors appeared in (\req(ww*k)),
\bea(ll)
\mu^{m+1} \bigg( \prod_{l=-m}^{m+1} (1- \omega^l {\tt x}) \bigg) {\tt W}_{p,\sigma}(m+1)&= 
\prod_{l=0}^m (1- \omega^{-l} {\tt x})({\tt y}- \omega^{-l} ),  \\ 
\mu^{m} \bigg( \prod_{l = -m}^{m-1} (1- \omega^l{\tt y})^{-1} \bigg)  \overline{\tt W}_{{\sf p},\sigma }(-m) &= \prod_{l =1}^{m} \frac{1}{(1- \omega^{-l}{\tt y})({\tt x}- \omega^{-l})} .
\elea(nf)
(The right side of the above second relation is defined to be $1$ when $m=0$.)
The normalized operators,
\bea(l)
\widetilde{Q}_R (\sigma) = \bigg( \prod_{l=0}^m (1- \omega^{-l} {\tt x})({\tt y}- \omega^{-l} ) \bigg)^L Q_R (\sigma) , \\ \widetilde{Q}_L (\sigma) = \bigg( \prod_{l =1}^{m} (1- \omega^{-l}{\tt y})({\tt x}- \omega^{-l}) \bigg)^{-L}  Q_L (\sigma) ,
\elea(NQRL)
are the trace of monodromy matrices using the normalized local ${\tt S}$, $\widehat{\tt S}$-matrices:
\bea(l)
S_{i, j}(\sigma)  =\bigg( \prod_{l=0}^m (1- \omega^{-l} {\tt x})({\tt y}- \omega^{-l} )\bigg) {\tt S}_{i, j} =  X^i \widetilde{w}(\sigma; j-i)\tau_{i, j} , \\
\widehat{S}_{i, j}(\sigma)  = \bigg(\prod_{l =1}^{m} \frac{1}{(1- \omega^{-l}{\tt y})({\tt x}- \omega^{-l})} \bigg) \widehat{\tt S}_{i, j}= 
\widehat{\tau}_{i, j} \widetilde{w^*} ( \sigma; j-i) X^{-i} ,
\elea(SShat*)
for $i, j \in \ZZ_N$ and $\sigma \in {\goth S}$, where by (\req(ww*k)) and (\req(nf)), the cyclic vectors $\widetilde{w} (\sigma; k), \widetilde{w^*} ( \sigma; k)$ are defined by 
$$
\widetilde{w} (\sigma; k)_n = \overline{\tt W}_{{\sf p}',\sigma}(n) {\tt W}_{{\sf p},\sigma}(n- k), \ \ 
\widetilde{w^*} ( \sigma; k)^n = 
\overline{\tt W}_{{\sf p},\sigma }(k -n ) {\tt W}_{{\sf p}',\sigma }( -n ) .
$$
We now examine the relation between $\widetilde{Q}_L (\sigma) \widetilde{Q}_R (\sigma')$ and $\widetilde{Q}_L (\sigma') \widetilde{Q}_R (\sigma)$ through the product function of vectors in $\widehat{S}_{i, j}, S_{k, l}$ (\req(SShat*)):
$$
\begin{array}{ll}
F (\sigma, \sigma' | i, j ; k, l) &= \widetilde{w^*}(\sigma; j-i) X^{-i+k} \widetilde{w}(\sigma'; l-k) \\
&= \sum_{n \in \ZZ_N} \overline{\tt W}_{{\sf p},\sigma }(j-n ) {\tt W}_{{\sf p}',\sigma }( i-n)  \overline{\tt W}_{{\sf p}',\sigma'}(n-k) {\tt W}_{{\sf p},\sigma'}(n-l).
\end{array}
$$
Parallel to the discussion of (\req(pfp)), the same argument leads to the constraint of $\sigma, \sigma'$ lying in a curve ${\goth W}_{k'}$ (\req(CPMc)) so that the star-triangle relation (\req(TArel)) holds for $({\sf p}, \sigma, \sigma')$ and $({\sf p}', \sigma, \sigma')$. This in turn yields the relation\footnote{The $F (\sigma, \sigma' | i, j ; k, l)$ is the same as $U_{qr}(a, b, c, d)$ in \cite{BBP} (2.28) where $p, p', q, r, a, b, c, d$ correspond to ${\sf p}', {\sf p}, \sigma, \sigma', i, j, l, k$ here. The derivation of (\req(WFW)) here can be obtained by revising the arguments in \cite{BBP} (2.18)-(2.31).}
\be
{\tt W}_{\sigma, \sigma'} ( i - k ) F (\sigma, \sigma' | i, j ; k, l) {\tt W}_{\sigma, \sigma'}( j- l )^{-1} = \frac{f_{{\sf p}\sigma } f_{{\sf p}'\sigma'} }{f_{{\sf p}\sigma'} f_{{\sf p}'\sigma } } F (\sigma', \sigma | i, j ; k, l)
\ele(WFW)
where the function $f_{qr}$ is defined in (\req(Rf)). This implies 
\be
\widetilde{Q}_L (\sigma) \widetilde{Q}_R (\sigma')= (\frac{f_{{\sf p}\sigma } f_{{\sf p}'\sigma'} }{f_{{\sf p}\sigma'} f_{{\sf p}'\sigma } } )^L \widetilde{Q}_L (\sigma') \widetilde{Q}_R (\sigma), ~ ~ \sigma, \sigma' \in {\goth W}_{k'} .
\ele(LQRp')
We define the $Q$-operator
\be
Q (\sigma) = \widetilde{Q}_L (\sigma_0)^{-1} \widetilde{Q}_L (\sigma)= (\frac{f_{{\sf p}\sigma } f_{{\sf p}'\sigma_0 } }{f_{{\sf p}\sigma_0} f_{{\sf p}'\sigma } } )^L  \widetilde{Q}_R (\sigma)\widetilde{Q}_R (\sigma_0)^{-1} ~ \ ~ ~ ( \sigma \in {\goth W}_{k'})
\ele(Qpp')
where $\sigma_0$ is an arbitrary normalized point at which both $\widetilde{Q}_L (\sigma_0), \widetilde{Q}_R (\sigma_0)$ are non-singular. Then $Q (\sigma)$ for $\sigma \in {\goth W}_{k'}$ form a family of commuting operators satisfying the $TQ$-relation
\bea(ll)
 \tau^{(2)}(\omega^{-1} {\tt t} ) Q (\sigma) 
= & \frac{(  1  -  \omega^{-m} {\tt t}  )^L( \omega^{1+m} {\tt x}- 1)^L}{(\omega{\tt x}- 1)^L}  Q ( U_1 \sigma) X  
\\
&+ \frac{(1 -  \omega^m {\tt t} )^L({\tt x}- 1)^L}{( \omega^m {\tt x}- 1)^L}  Q (U_1^{-1} \sigma) .
\elea(TQpp')
As in (\req(Sdef)) by setting $\tau_{i, j} = \langle j |$, $ \widehat{\tau}_{i, j}= | j \rangle$ in (\req(SShat*)), one can identify $\widetilde{Q}_R, \widetilde{Q}_L$ with the CPM transfer matrices $T_{{\sf p}, {\sf p}'}, \widehat{T}_{{\sf p}, {\sf p}'}$ for the two vertical rapidities ${\sf p}, {\sf p}'$ in (\req(p'p)), $\widetilde{Q}_R (\sigma ) = T_{{\sf p}, {\sf p}'}(\sigma), \widetilde{Q}_R (\sigma ) = \widehat{T}_{{\sf p}, {\sf p}'}(\sigma)$ (\cite{BBP} (2.15a) and (2.15b)):
\bea(l)
\langle j_1, \ldots, j_L| \widetilde{Q}_R (\sigma ) |j'_1, \ldots, j'_L \rangle  = \prod_{\ell=1}^L  {\tt W}_{{\sf p}, \sigma} (j_\ell-j'_\ell) \overline{\tt W}_{{\sf p}', \sigma}(j_{\ell+1}-j'_\ell) ; \\
\langle j_1, \ldots, j_L|  \widetilde{Q}_L (\sigma) |j'_1, \ldots, j'_L \rangle = 
\prod_{\ell=1}^L 
\overline{\tt W}_{{\sf p}, \sigma} (j_{\ell}-j'_\ell) {\tt W}_{{\sf p}' \sigma} (j_\ell -j'_{\ell+1`}) .
\elea(QTpp')
Hence we have shown the following result.
\begin{thm}\label{thm:CPMpp} 
Let $\tau^{(2)}({\tt t})$ be the $\tau^{(2)}$-matrix $(\req(tau2))$ for the $L$-operator $(\req(Ktau))$ with $\varsigma= \omega^m ~ ( m \geq 0)$, and $T_{{\sf p}, {\sf p}'}, \widehat{T}_{{\sf p}, {\sf p}'}$ be the CPM transfer matrices for two vertical superintegrable rapidities ${\sf p}, {\sf p}'$ in $(\req(p'p))$. Then the $Q_R, Q_L$-operators of the $\tau^{(2)}$-model are 
$$
\begin{array}{l}
Q_R (\sigma ) = \bigg( \prod_{l=0}^m (1- \omega^{-l} {\tt x})({\tt y}- \omega^{-l} )\bigg)^{-L} T_{{\sf p}, {\sf p}'}( \sigma) , \\
 Q_L(\sigma )= \bigg(\prod_{l =1}^{m} (1- \omega^{-l}{\tt y})({\tt x}- \omega^{-l}) \bigg)^L \widehat{T}_{{\sf p},{\sf p}'}( \sigma)
\end{array}
$$
for $\sigma \in {\goth S}$, which satisfy the $TQ$-relation $(\req(TQa))$.
The commuting relation $(\req(LQRp'))$ for the normalized $\widetilde{Q}_R, \widetilde{Q}_L$-operators in $(\req(NQRL))$, i.e. $T_{{\sf p}, {\sf p}'}$ and $\widehat{T}_{{\sf p},{\sf p}'}$,  holds when both $\sigma, \sigma'$ are in a CPM curve ${\goth W}_{k'}$, and  the $Q$-operator defined by in $(\req(Qpp'))$ satisfies the $TQ$-relation $(\req(TQpp'))$. 
\end{thm}
\par \noindent 
{\bf Remark}. For odd $N=2M+1$, by Theorem \ref{thm:crepT} the $\tau^{(2)}$-models in the above theorem are equivalent to the XXZ chain
$T(s)$ in (\req(TcXZ)) associated to the cyclic $U_{\sf q}(sl_2)$-representations $\rho_\varepsilon$  with $\varepsilon \in \ZZ$. As a consequence, the theory of the XXZ chain for a cyclic representation of $U_{\sf q}(sl_2)$ with the parameter ${\sf q}^{\varepsilon N}=1$ can be identified with the inhomogeneous CPM with two vertical superintegrable rapidities ${\sf p}, {\sf p}'$. In particular when $\varsigma =  \omega^M$ (equivalently $\varepsilon = M$),  ${\sf p}= {\sf p}'$ in (\req(p'p)), which is equal to the superintegrable element (\req(supp)) with $({\rm m}, c)= (M, 0)$. This again shows the identical theory between the spin-$\frac{N-1}{2}$ XXZ chain and the homogeneous superintegrable CPM at ${\sf p}$ as described in Theorem \ref{thm:crepT}. Indeed in this case,  the cyclic vectors $w, w^*$ in (\req(ww*k)) and $v, v^*$ (\req(vv*Gp)) are related by 
$$
 \omega^{\frac{1}{2}} (1-{\tt x}^N)  w( \sigma; k) =  v ( \sigma ; k), ~    \\
 \omega^{-\frac{1}{2}}  ({\tt y}^N -1)^{-1}   w^{*} ( \sigma; k) =v^{* }(\sigma; k) , 
$$
by which, arguments in Theorems \ref{thm:CPMpp} and \ref{thm:CPQp} are equivalent.

\section{Concluding Remarks}\label{sec. F} 
In this article, we have successfully constructed the superintegrable $N$-state CPM transfer matrix as the $Q$-operator of the XXZ chain for cyclic $U_{\sf q} (sl_2)$-representations with $N$th root-of-unity property for odd $N$. By converting the root-of-unity XXZ chain with cyclic representations of $U_{\sf q} (sl_2)$ for ${\sf q}^N=1$ to a special one-parameter family of generalized $\tau^{(2)}$-models (\req(Ktau)), we construct the $Q_R, Q_L$ and $Q$-operators of those $N$th-root-of-unity $\tau^{(2)}$-models by the Baxter's method of producing the $Q_{72}$-operator in the root-of-unity eight-vertex model \cite{B72}. With a correct identification of parameters in the $Q$-operator construction from the $\tau^{(2)}$-model, the Boltzmann weights of CPM are found to express the $Q_R, Q_L$-operators, which are identified with transfer matrices in the theory of  superintegrable CPM with two vertical rapidities. We also apply the techniques to the superintegrable $\tau^{(2)}$-models (\req(SUL)), thus obtain the transfer matrices of homogeneous superintegrable CPM as the $Q_R, Q_L$-operators. We describe the steps in some detail for the special superintegrable $\tau^{(2)}$-model (\req(SupL)) in section \ref{ssec.Qsup1} as it will serve as a model example of constructing $Q$-operator for the symmetry study of other lattice models. As a result of our working, the spin-$\frac{N-1}{2}$ XXZ  chain model and the superintegrable CPM are unified into one single theory for odd $N$ (Theorem \ref{thm:crepT}), which provides a satisfactory connection between these two models in both qualitative and quantitative aspects. Further possible extension to root-of-unity XXZ spin chain of other higher spins seems somewhat subtle, but the work is under consideration.
For the $Q$-operators in section \ref{ssec.IcpQ}, the CPM transfer matrices we obtain here is an inhomogeneous one with two vertical rapidities, a significant difference from those homogenous CPM in section \ref{sec:Qtau}. The finding suggests that there should be a relationship between inhomogeneous CPM theory and the generalized $\tau^{(2)}$-models. The connection is not immediately apparent, and  much remains to be discovered in this direction.

\section*{Acknowledgements}
The author wishes to thank Professor I. Satake for the invitation of visiting U. C. Berkeley in the spring of 2007,  where part of this work was carried out.
This work is supported in part by National Science Council of Taiwan under Grant No NSC 95-2115-M-001-007.


\begin{thebibliography}{99}
\bibitem{AMP} G. Albertini, B. M. McCoy, and 
J. H. H. Perk, Eigenvalue spectrum of the
superintegrable chiral Potts model, In {\it Integrable system in quantum field theory and statistical mechanics,}  Adv. Stud. Pure Math., 19, Kinokuniya Academic, Academic Press, Boston, MA (1989) 1--55.
%
\bibitem{AMPTY} H. Au-Yang, B. M. McCoy,  
J. H. H. Perk, S. Tang and M. L. Yan, Commuting transfer matrices in chiral Potts models: solutions of the star-triangle equations with genus $> 1$, Phys. Lett. A 123 (1987) 219--223. 
%
\bibitem{B72} R. J. Baxter, Partition function of the eight vertex model, Ann. Phys. 70 (1972) 193--228.
%
\bibitem{Bax} R. J. Baxter, Exactly solved models
in statistical mechanics, Academic Press (1982).
%
\bibitem{B90} R. J. Baxter, Chiral Potts model: eigenvalues of the transfer matrix, Phys. Lett. A 146 (1990) 110--114.
% 
\bibitem{B93} R. J. Baxter, Chiral Potts model with skewed boundary conditions, J.
Stat. Phys. 73 (1993) 461--495.
%
\bibitem{B04} R. J. Baxter, The six and eight-vertex models revisited, J. Stat. Phys. 114 (2004) 43--66; cond-mat/0403138.
%
\bibitem{B05o} R. J. Baxter, The order parameter of the chiral Potts model, J. Stat. Phys. 120 (2005) 1--36; cond-mat/0501226.
%
\bibitem{B05} R. J. Baxter, Derivation of the order parameter of the chiral Potts model, Phys. Rev. Lett. 94 (2005) 130602; cond-mat/0501227.
%
\bibitem{BBP} R. J. Baxter, V.V. Bazhanov and
J.H.H. Perk,  Functional relations for transfer
matrices of the chiral Potts model, Int. J. Mod.
Phys. B 4 (1990) 803--870.
%
\bibitem{BPA} R. J. Baxter, J. H. H. Perk and H. Au-Yang, New solutions of the  star-triangle relations for the chiral Potts model, Phys. Lett. A 128 (1988) 138--142.
%
\bibitem{BazS} V.V. Bazhanov and Yu.G. Stroganov, Chiral
Potts model as a descendant of the six-vertex model, J.
Stat. Phys. 59 (1990) 799--817.
%
\bibitem{DKM} S. Dasmahapatra, R. Kedem and B. M. McCoy, Physics beyond quasi-particles: Spectrum and completeness of the 3 state superintegrable chiral Potts model, Nucl. Phys. B 396 (1993) 506--540; hep-th/9204003.
%
\bibitem{DJMM} E. Date, M. Jimbo, K. Miki and T. Miwa, Cyclic representations of $U_q(sl(n+1, \CZ))$ at $q^N=1$, Publ. RIMS, Kyoto Univ. 27 (1991) 347--366.
%
\bibitem{DFM} T. Deguchi, K. Fabricius and B. M. McCoy, The $sl_2$ loop algebra symmetry for the six-vertex model at roots of unity, J. Stat. Phys. 102 (2001) 701--736; cond-mat/9912141. 
%
\bibitem{De01a} T. Deguchi: Construction of some missing eigenvectors of the XYZ spin chain at the discrete coupling constants and the exponentially large spectral degeneracy of the transfer matrix, J. Phys. A: Math. Gen. 35 (2002) 879--895; cond-mat/0109078. 
%
\bibitem{De01b} T. Deguchi: The 8V CSOS model and the $sl_2$ loop algebra symmetry of the six-vertex model at roots of unity, Int. J. Mod. Phys. B 16 (2002) 1899--1905; cond-mat/0110121. 
%
\bibitem{De05} T. Deguchi: Regular XXZ Bethe states at roots of unity- as highest weight vectors of the $sl_2$ loop algebra at roots of unity, cond-mat/0503564 v3. 
%
\bibitem{FM00} K. Fabricius and B. M. McCoy, Bethe's equation is incomplete for the XXZ model at roots of unity, J. Stat. Phys. 103 (2001) 647--678; cond-mat/0009279.
%
\bibitem{FM001} K. Fabricius and B. M. McCoy, Completing Bethe equations at roots of unity, J. Stat. Phys. 104 (2001) 573--587; cond-mat/0012501.
%
\bibitem{FM01} K. Fabricius and B. M. McCoy, Evaluation parameters and Bethe roots for the six vertex model at roots of unity, {\it Progress in Mathematical Physics} Vol 23, eds. M. Kashiwara and T. Miwa,  Birkh\"{a}user Boston (2002), 119--144; cond-mat/0108057.
%
\bibitem{FM02} K. Fabricius and B. M. McCoy, New developments in the eight vertex model, J. Stat. Phys. 111 (2003) 323--337; cond-mat/0207177.
%
\bibitem{FM04} K. Fabricius and B. M. McCoy, Functional equations and fusion matrices for the eight vertex model, Publ. RIMS, 40 (2004) 905--932; cond-mat/0311122.
%
\bibitem{FM41} K. Fabricius and B. M. McCoy, Root of unity symmetries in the 8 and 6 vertex models, cond-mat/0411419.
%
\bibitem{F06} K. Fabricius, A new $Q$-operator in the eight-vertex model, cond-mat/0610481v3.
%
\bibitem{Fad} L. D. Faddeev, How algebraic Bethe
Ansatz works for integrable models, eds. A.
Connes, K. Gawedzki and J. Zinn-Justin, {\it
Quantum symmetries/ Symmetries quantiques},
Proceedings of the Les Houches summer school,
Session LXIV, Les Houches, France, August 1-
September 8, 1995, North-Holland (1998),  149--219.
%
\bibitem{GR} G. von Gehlen and R. Rittenberg,   
$Z_n$-symmetric quantum chains with infinite set of
conserved charges and $Z_n$ zero modes, Nucl. Phys. B 257
(1985) 351--370.
%
\bibitem{HKN} S. Howes, L.P. Kadanoff and M. den
Nijs, Quantum model
for commensurate-incommensurate transitions, Nucl.
Phys. B 215 (1983) 169--208.
%
\bibitem{KiR} A, N. Kirillov and N. Yu. Reshetikhin, Exact solution of the integrable XXZ Heisenberg model with arbitrary spin: I. The ground state and the excitation spectrum, J. Phys. A: Math. Gen. 20 (1987) 1565 -- 1595.
%
\bibitem{KBI} V. E. Korepin, N. M. Bogoliubov, and A. G. Izegin, Quantum inverse scattering method and correlation functions, Cambridge Univ. Press, Cambridge, 1993.
%
\bibitem{KRS} P. P. Kulish,  N. Yu. Reshetikhin and E. K. Sklyanin, Yang Baxter equation and representation theory, Lett. Math. Phys. 5 (1981) 393--403.
%
\bibitem{KS} P. P. Kulish and E. K. Sklyanin,
Quantum spectral transform method. Recent
developments, eds. J. Hietarinta and C. Montonen,
Lecture Notes in Physics 151 Springer (1982),
61--119.
%
\bibitem{MPTS} B. M. McCoy,  
J. H. H. Perk, S. Tang and C. H. Sah, Commuting transfer matrices for the four-state self-dual chiral Potts model with a genus-three uniformizing Fermat curve, Phys. Lett. A 125 (1987) 9--14.
%
\bibitem{MR} B. M. McCoy and S. S. Roan, Excitation spectrum and phase structure of the chiral Potts model. Phys. Lett. A 150 (1990) 347--354.
%
\bibitem{NiD} A. Nishino and T. Deguchi, The $L(sl_2)$ symmetry of the Bazhanov-Stroganov model associated with the superintegrable chiral Potts model, Phys. Lett. A 356 (2006) 366--370
; cond-mat/0605551.
%
\bibitem{On} L. Onsager, Crystal statistics. I. A two-dimensional model with an order-disorder transition, Phys. Rev. 65 (1944) 117--149. 
%
\bibitem{R91} S. S. Roan, Onsager's algebra, loop algebra and chiral Potts model, Preprint MPI/91-70 
Max-Planck-Inst. fur Math., Bonn,  1991.
%
\bibitem{R04} S. S. Roan, Chiral Potts rapidity curve descended from six-vertex model and symmetry group of rapidities, J. Phys. A: Math. Gen. 38 (2005) 7483--7499; cond-mat/0410011.
%
\bibitem{R05o} S. S. Roan, The Onsager algebra symmetry of $\tau^{(j)}$-matrices in the superintegrable chiral Potts model, J. Stat. Mech. (2005) P09007; cond-mat/0505698.
%
\bibitem{R05b} S. S. Roan, Bethe ansatz and symmetry in superintegrable chiral Potts model and root-of-unity six-vertex model, in Nankai Tracts in Mathematics Vol. 10, {\it Differential Geometry and Physics}, eds. Mo-Lin Go and Weiping Zhang,  World Scientific, Singapore (2006), 399-409; cond-mat/0511543.
%
\bibitem{R06Q} S. S. Roan, The Q-operator for root-of-unity symmetry in six vertex model, J. Phys. A: Math. Gen. 39 (2006) 12303-12325; cond-mat/0602375.
%
\bibitem{R06F} S. S. Roan, Fusion operators in the generalized $\tau^{(2)}$-model and root-of-unity symmetry of the XXZ spin chain of higher spin, J. Phys. A: Math. Theor. 40 (2007) 1481-1511; cond-mat/0607258.
%
\bibitem{R06Q8} S. S. Roan, The $Q$-operator and functional relations of the eight-vertex model at root-of-unity $\eta = 2mK/N$ for odd $N$, J. Phys. A: Math. Theor. 40 (2007) 11019-11044; cond-mat/0611316.
%
\bibitem{R07} S. S. Roan, On Q-operators of XXZ spin chain of higher spin, cond-mat/ 0702271.
\end{thebibliography}
\end{document}